\documentclass[twocolumn]{aastex63}

\usepackage{xspace}
\usepackage{threeparttable}
\usepackage{threeparttablex}
\usepackage{amsmath}

\newcommand{\solarmass}{M\ensuremath{_{\odot}}\xspace}

\newcommand{\coldenunit}{M$_{\odot}$ pc$^{-2}$\xspace}
\newcommand{\alphaco}{$\alpha_{\mathrm{CO}}$\xspace}
\newcommand{\alphacou}{$\mathrm{M_{\odot}\ (K\ km\ s^{-1}\ pc^{2})^{-1}}$\xspace}

\newcommand{\cotwo}{$^{12}$CO $J$=2-1\xspace}

\newcommand{\cothree}{$^{12}$CO $J$=3-2\xspace}

\shorttitle{YMCs in the Antennae}
\shortauthors{He et al.}
\begin{document}

\title{Embedded Young Massive Star Clusters in the Antennae Merger}

\correspondingauthor{Hao He}
\email{heh15@mcmaster.ca}

\author[0000-0001-9020-1858]{Hao He}
\affiliation{McMaster University \\
1280 Main St W, Hamilton, ON L8S 4L8, CAN}

\author{Christine Wilson}
\affiliation{McMaster University \\
	1280 Main St W, Hamilton, ON L8S 4L8, CAN}

\author{Nathan Brunetti}
\affiliation{McMaster University \\
	1280 Main St W, Hamilton, ON L8S 4L8, CAN}

\author{Molly Finn}
\affiliation{University of Virginia,  \\
	530 McCormick Road, Charlottesville, VA 22904, US}

\author{Ashley Bemis}
\affiliation{Leiden University, \\
	 Rapenburg 70, 2311 EZ Leiden, Netherlands}

\author{Kelsey Johnson}
\affiliation{University of Virginia,  \\
	530 McCormick Road, Charlottesville, VA 22904, US}
\affiliation{Adjunct Astronomer at the National Radio Astronomy Observatory}



\begin{abstract}

The properties of young massive clusters (YMCs) are key to understanding the star formation mechanism in starburst systems, especially mergers. We present ALMA high-resolution ($\sim$10 pc) continuum (100 and 345 GHz) data of YMCs in the overlap region of the Antennae galaxy. We identify 6 sources in the overlap region, including two sources that lie in the same giant molecular cloud (GMC). These YMCs correspond well with radio sources in lower resolution continuum (100 and 220 GHz) images at GMC scales ($\sim$60 pc). We find most of these YMCs are bound clusters through virial analysis. We estimate their ages to be $\sim$1 Myr and to be either embedded or just beginning to emerge from their parent cloud. We also compare each radio source with Pa$\beta$ source and find they have consistent total ionizing photon numbers, which indicates they are tracing the same physical source. By comparing the free-free emission at $\sim$10 pc scale and $\sim$60 pc scale, we find that $\sim$50\% of the free-free emission in GMCs actually comes from these YMCs. This indicates that roughly half of the stars in massive GMCs are formed in bound clusters. We further explore the mass correlation between YMCs and GMCs in the Antennae and find it generally agrees with the predictions of the star cluster simulations. The most massive YMC has a stellar mass that is 1\% -- 5\% of its host GMC mass. 
%We find no evidence that feedback from these YMCs increases the GMC temperature and thus no evidence that feedback from YMCs has changed the global properties of GMCs. 
\end{abstract}



\section{Introduction} \label{sec:intro}

How massive star clusters form is one of the major unsolved problems in star formation. Massive cluster formation was ubiquitous in the early universe, as witnessed by the populations of old massive globular clusters found in galaxies of all masses and morphologies \citep{Harris_2013}. Indeed, given that the fraction of stellar mass contained in globular clusters today may be 10\% or less of their initial mass as protoclusters \citep{Fall_2001,Whitmore_2007,Li_2014}, massive clusters should have been one of the most important modes of star formation in the early universe. In addition, current theory suggests that star formation is caused by fragmentation of hierarchically collapsing giant
molecular clouds \citep[GMCs; e.g.][]{McKee_Ostriker_2007}, which naturally leads to the conclusion that stars tend to form together in bound clusters. Both theory \citep{Kruijssen_2012} and observations \citep{Adamo_2020} suggest that for ultra/luminous infrared galaxies (U/LIRGs), more than 50\% of stars are formed in bound clusters. Therefore, studying young massive star clusters (YMCs) will help us understand the star forming process in starburst systems. 

%the cluster formation efficiency (CFE), which quantifies the fraction of stars formed in bound clusters, is dependent on the star formation rate surface density, $\Sigma_{\mathrm{SFR}}$ (Fig. 1). For starburst systems such as NGC 3256, more than 60\% of stars are expected to form in star clusters. Therefore, studying young massive star clusters (YMCs) will help us to understand the star formation process in extreme starburst systems in the local universe.

Large populations of young massive star clusters are seen in a diverse range of interacting galaxies and merger remnants by the Hubble Space Telescope (HST), from the M51 system \citep{Scoville_2001} to the Antennae \citep{Whitmore_1999, Whitmore_2010} to Arp 220 \citep{Scoville_2000, Zhang_2001}. Of all the systems studied so far, the Antennae stands out for its uniquely large population of young massive clusters \citep{Scoville_2000, Wilson_2006} and massive molecular clouds \citep{Wilson_2003}, while its proximity \citep[22 Mpc][]{Schweizer_2008} allows us to obtain the highest possible spatial resolution. Multi-wavelength observations have mapped out the distribution of optically visible young clusters \citep{Whitmore_2010} as well as the far-infrared emission that traces buried star formation \citep{Klaas_2010}. Much of the far-infrared emission is located in the “overlap region” \citep{Stanford_1990}, a region that is also rich in molecular gas \citep{Wilson_2003, Whitmore_2014, Schirm_2016} and radio continuum emission \citep{Neff_2000}. The overlap region also contains two bright water masers, which are a common indicator of massive star formation \citep{Brogan_2010}. However, optical observations generally miss extremely young clusters \citep[ages $<$ few Myr, ][]{Johnson_2003,Johnson_2004, Reines_2008, Johnson_2009, Hannon_2019} that have high dust extinction. \citet{Whitmore_2010} suggest that about 16\% of star clusters in the Antennae are hidden from view in the optical. However, since radio observations are generally unaffected by dust extinction \citep{Murphy_2011}, we can use radio frequencies to probe these extremely young YMCs. 

In this paper, we measure YMC properties using ALMA continuum images at $\sim$ 10 pc scale. We compare these images with continuum images at GMC scales ($\sim$60 pc) to explore various correlations between the YMCs and their host GMCs. In Section 2, we describe the observations and how we processed the data. In Section 3, we describe how we measure various quantities, such as free-free flux, dust flux, temperature and velocity dispersion. We then derive the stellar mass and gas mass based on those quantities. In Section 4, we use those quantities to explore the evolutionary stage and dynamical state of the YMCs. In Section 5, we compare various quantities, such as star formation rate (SFR) and total mass at YMC and GMC scales, to study the correlation between these two types of objects.

\section{Observations and Data Reduction} \label{sec:method}

\begin{table*} % different from the paper version. 
	\centering
	\caption{Summary of the ALMA continuum observations of the Antennae}
	\label{tab:observation}
	\begin{threeparttable}
	\begin{tabular}{lccccc}
		\hline
		Project Code   & Central         & Beam              & Arrays used        & LAS $^a$  & RMS noise         \\
		& Frequency (GHz) &    (")                  &              &     (")    & (mJy beam$^{-1}$) \\ \hline
		2018.1.00272.S & 100             & 0.57 $\times$ 0.43   & 12m+7m  & 70      & $0.011 $          \\
		2018.1.00272.S & 220             & 0.63 $\times$ 0.59   & 12m+7m  & 41      & $0.054 $          \\
		2016.1.00041.S & 100             & 0.11 $\times$ 0.11 & 12m     & 4.1     & $0.016$           \\
		2016.1.00041.S & 345             & 0.11 $\times$ 0.11  & 12m     & 4.1     & $0.04$           \\ \hline
	\end{tabular}
	\begin{tablenotes}
		\item a. LAS stands for largest angular scale. 
	\end{tablenotes}
	\end{threeparttable}
\end{table*}

\begin{figure*}
	\epsscale{1.1}
	\plottwo{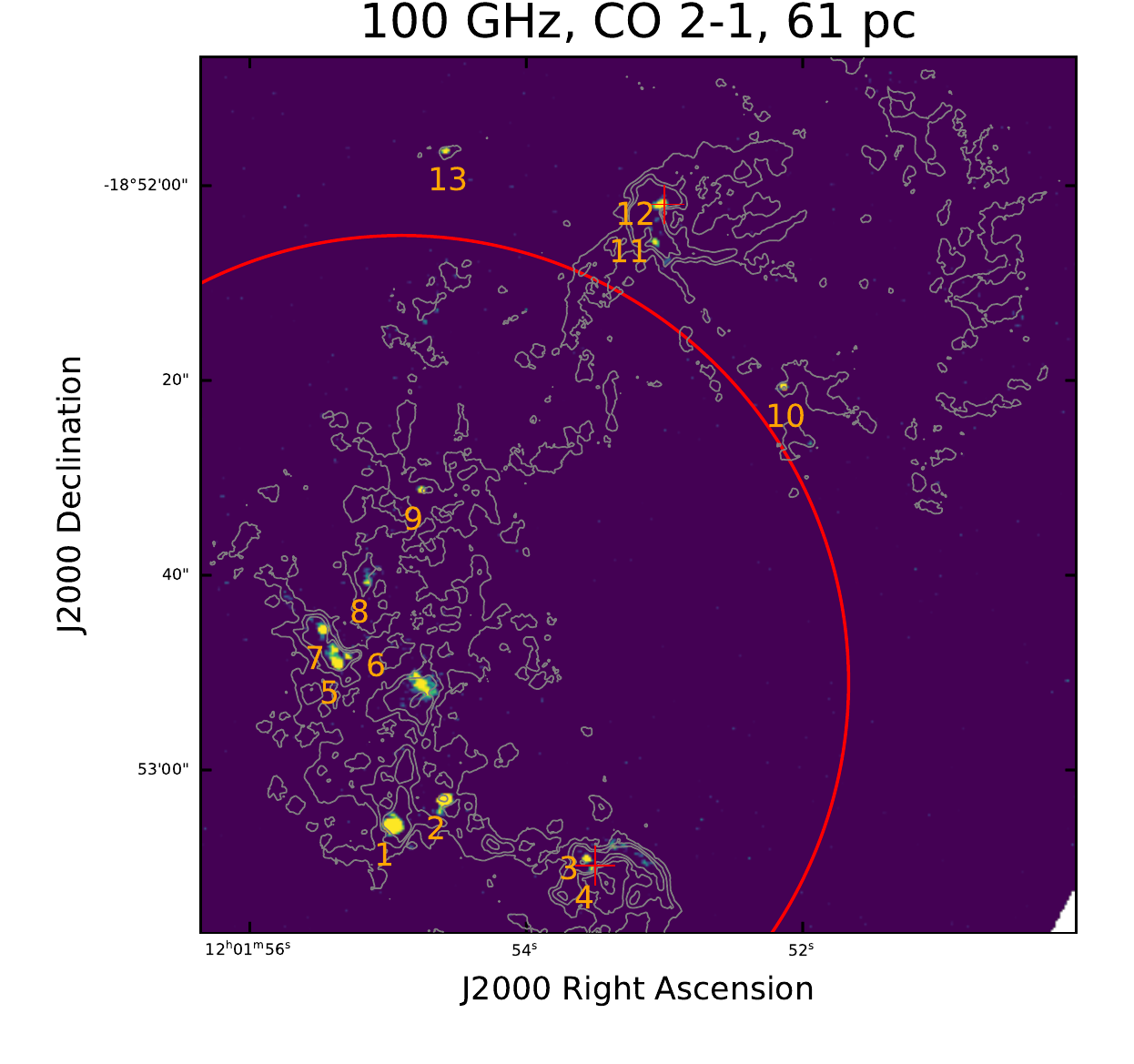}{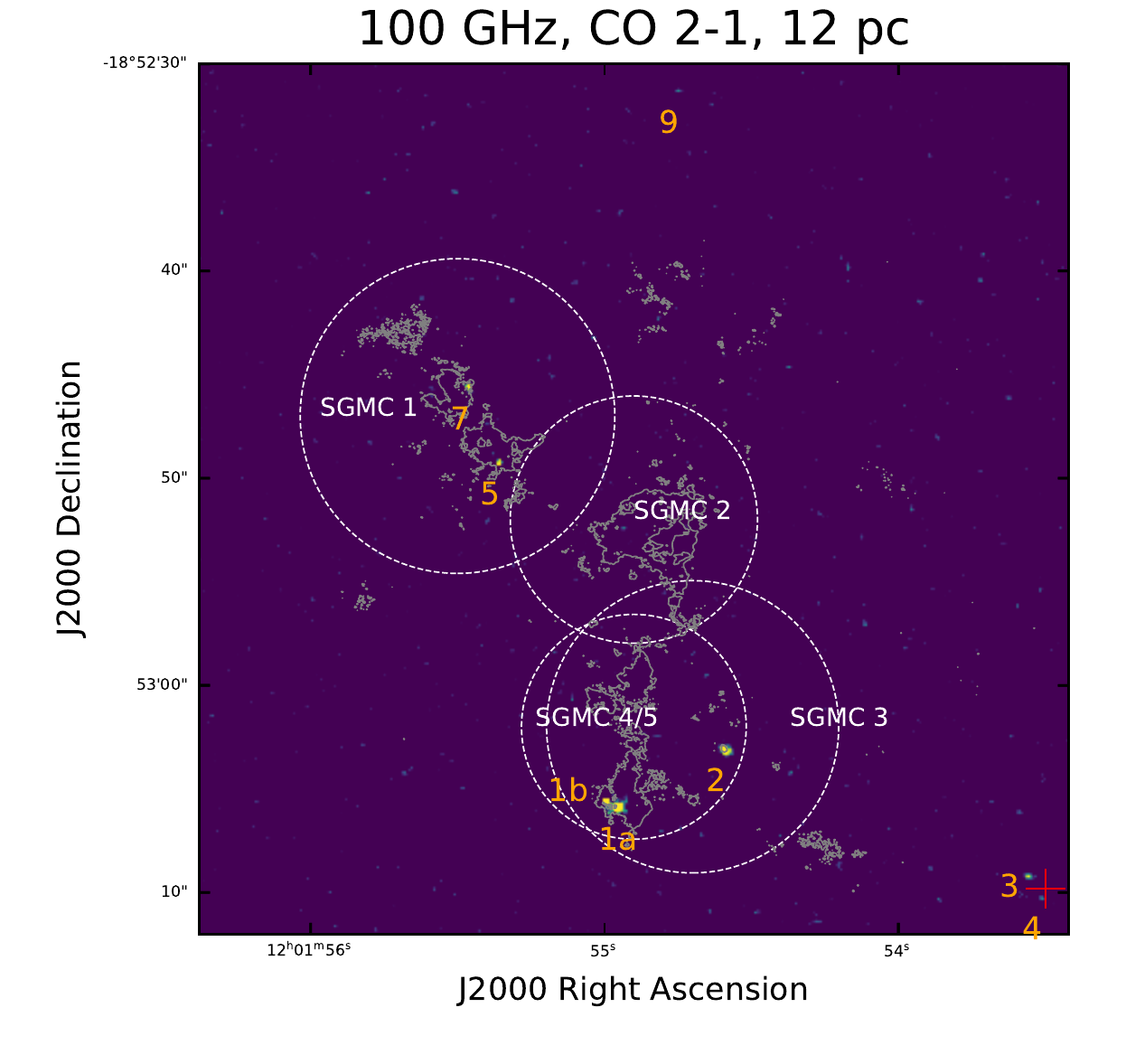}
	\plottwo{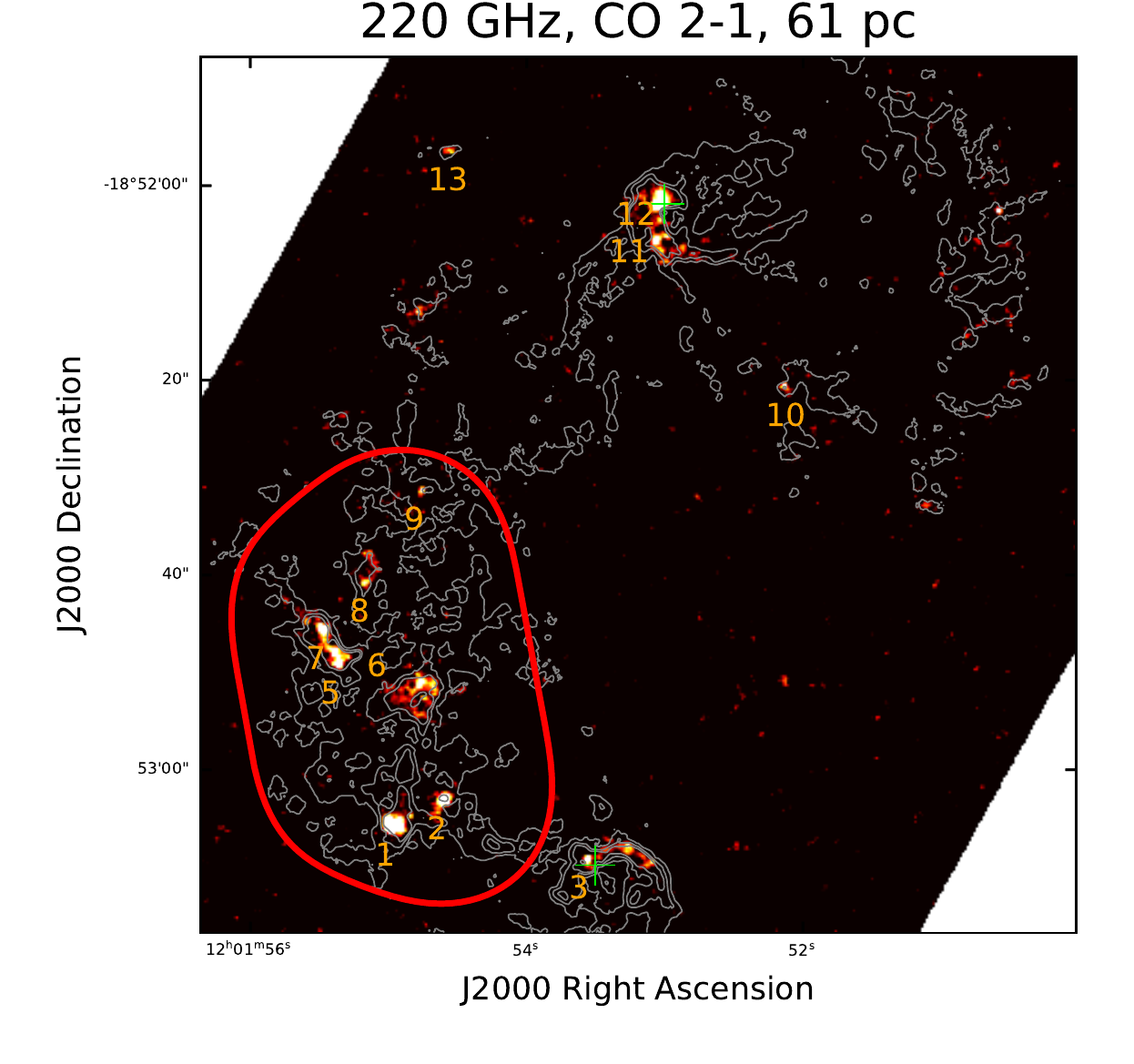}{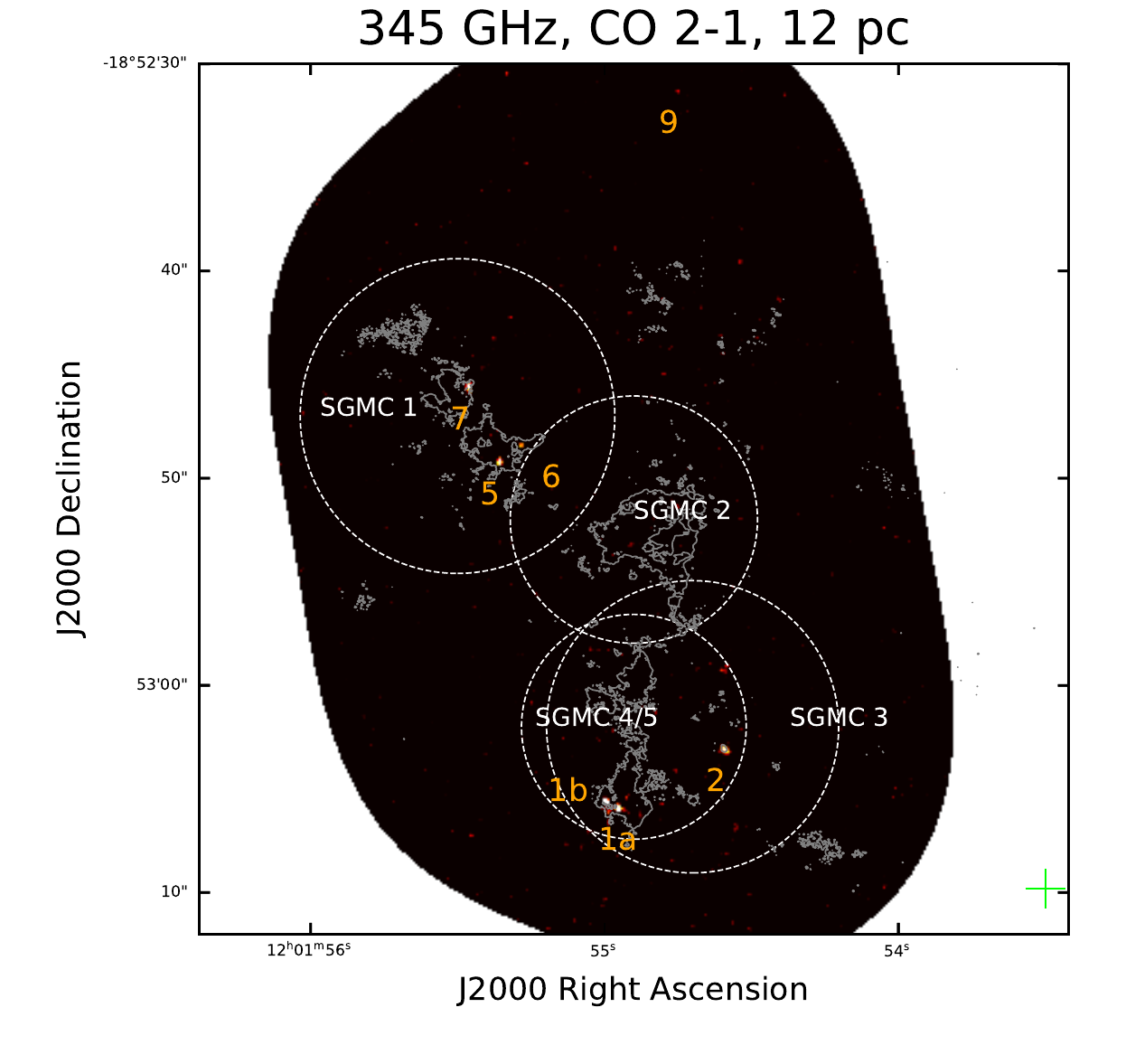}
	\caption{Continuum image of the Antennae at (Upper) 100 GHz, (Lower Left) 230 GHz and (Lower Right) 340 GHz. The continuum images in the left panels are from ALMA project 2018.1.00272.S with physical resolution of 61 pc. The contours are from the $^{12}$CO J=2-1 moment 0 map from Brunetti et al. (in prep). Red apertures show the field of view of the high-resolution images shown in the right panels. The continuum images in the right panels are from ALMA project 2016.1.00041.S with physical resolution of 12 pc. Contours in the right panels are from the \cotwo moment 0 map from \citet{Finn_2019}. The dashed circles in the right panels show the location of the SGMCs identified in \citet{Wilson_2000} with the diameters equal to those of the SGMCs. The red and green plus signs show the locations of two nuclei from \citet{Zhang_2001}.  }
	\label{fig:cont_image}
\end{figure*}

\begin{figure*}
	\plotone{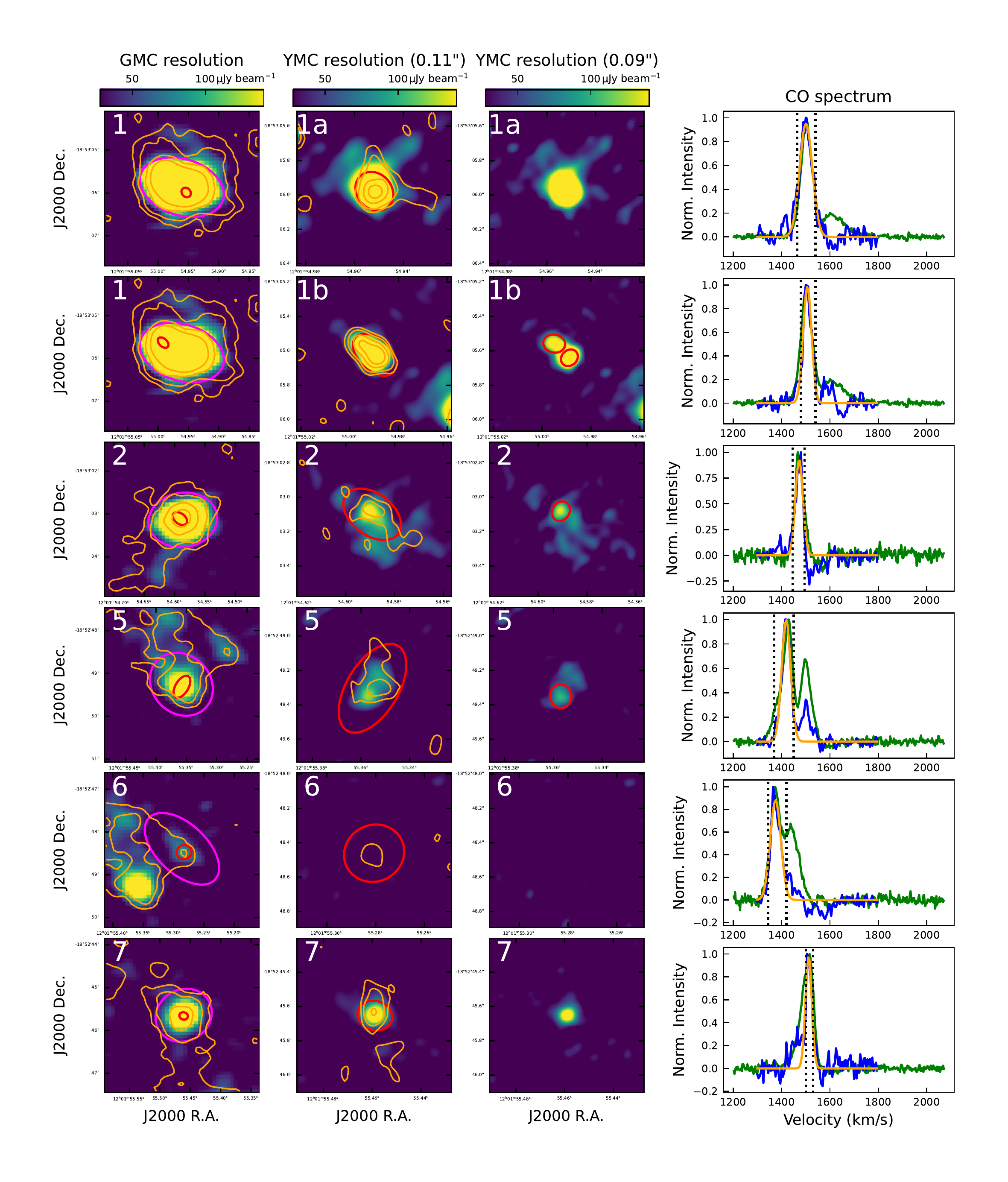}
	\caption{(Left) Continuum images and (Right) \cotwo spectra of individual YMCs. For the images, the left column shows the 100 GHz data at GMC resolution (61 pc). The orange contours are the 220 GHz data at similar resolution. Magenta apertures are the apertures used to measure the flux at GMC scales. Red apertures are the apertures used to measure the flux at YMC scales. The middle and right column show the 100 GHz data at YMC scales with slightly different resolution. The orange contours in the middle column are the 345 GHz data at the same resolution. Red apertures in the middle column are the same as the ones in the left column. The red apertures in the right column are the fitted Gaussian beam for substructures in some of the YMCs. The \cotwo spectra are normalized to the peak of the line emission measured in the magenta apertures in the left column (green spectrum )and the red apertures in the middle column (blue spectrum). The orange curves are the fitted Gaussian function to the measured spectrum (shown in blue). The vertical dotted lines specify the velocity ranges that we use to fit the Gaussian spectra.  }
	\label{fig:cluster_summary}
\end{figure*}

\begin{figure*}
	\plotone{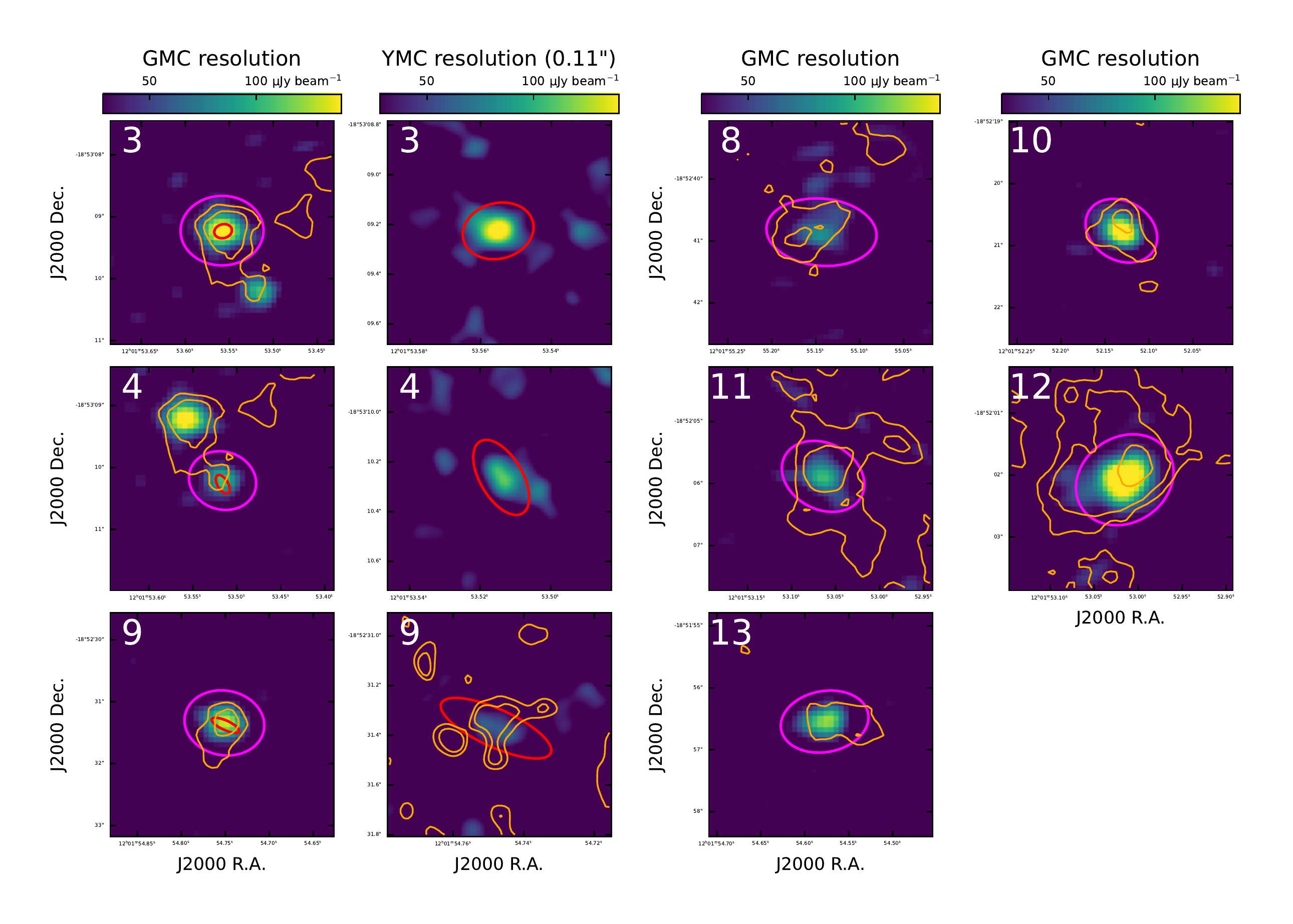}
	\caption{Images for YMCs that do not have \cotwo observations at 12 pc scale. (Left) the 3 sources that have 100 GHz images at YMC scale. (Right) The remaining sources that only have 100 GHz GMC resolution data. The orange contour in GMC-resolution images are the 220 GHz continuum and that in YMC-resolution images are the 345 GHz continuum. The magenta apertures are used to measure the GMC fluxes while the red apertures are used to measure the YMC fluxes. }
	\label{fig:cluster_summary_appendix}
\end{figure*}

\subsection{Continuum at star cluster resolution}

The continuum data at star cluster resolution is from ALMA project 2016.1.00041.S (PI: Christine Wilson). This project only has continuum observations from the 12m array and covers ALMA Bands 3 (100 GHz) and 7 (345 GHz). The total bandwidth for each spectral window is 2000 MHz. The spatial resolution for both frequency bands is about 0.1 arcsec ($\sim$ 10 pc). The largest angular scale is about 4.1 arcsec ($\sim$ 500 pc, Table 1)

The original reduction scripts were used to calibrate the raw data using CASA version 4.7.2. All of the imaging steps were carried out using CASA version 5.1.0-70. Before imaging, we binned channels in the calibrated measurement set to a channel width of 125 MHz for Band 3 and 250 MHz for Band 7. We then ran \texttt{tclean} on the measurement set and did the clean interactively. To match the spatial scale of the Band 3 and Band 7 images, we set the lowest uvrange to be 50 k$\lambda$. We used a robust parameter of 0.5 for the \texttt{tclean} command. Finally, we used the CASA command \texttt{imsmooth} to smooth both images to a beam size of 0.11" $\times$ 0.11" (12 pc).

\subsection{Continuum at GMC resolution}

The continuum data at GMC resolution is acquired from project 2018.1.00272.S (PI: Christine Wilson). This project has observations with both the 7m array and 12m extended and compact configuration arrays and covers frequencies in ALMA Bands 3 (110 GHz) and 6 (220 GHz) with various spectral lines detected (Brunetti et al. in prep.). The total usable bandwidth for each of the spectral windows is 1875 MHz for the 12m array and 2000 MHz for the 7m array. The highest spectral resolution is 0.976 MHz (Table 1). 

The original reduction scripts were used to calibrate the raw data using CASA version 5.4.0-70. We used CASA 5.6.1 to image the continuum in Band 3 and Band 6 using all of the spectral windows in each band. Before imaging, we flagged the channels with detected spectral lines and combined data from different arrays into a single measurement set. We also binned the channels to a channel width of 125 MHz to speed up the imaging process. Imaging used the CASA command \texttt{tclean} with the clean threshold set to be 2 times the RMS noise. We chose the specmode to be "mfs" and masking method to be 'auto-multithresh' \citep{Kepley_2020} to choose the clean regions automatically. The 'auto-multithresh'  parameters were left as the default values. We applied the primary beam correction to the cleaned images.

\subsection{CO line data at GMC resolution}
The \cotwo data with GMC resolution is also from project 2018.1.00272.S (PI: Christine Wilson). The image cube was made with a modified version of PHANGS pipeline \citep[; see details in Brunetti et al. in prep.]{Leroy_2021}. The beam size was rounded to 0.51" $\times$ 0.51" (54 pc). The velocity resolution is 2.65 km s$^{-1}$. The RMS is 2.6 K.

\subsection{CO line data at star cluster resolution}

The \cotwo data with star cluster resolution is from \citet{Finn_2019}. The beam size of the data is 0.12" $\times$ 0.09" (13 pc). The channel width of the cube is 5 km s$^{-1}$. The RMS of the cube is 1.2 mJy beam$^{-1}$. 
%The moment 0 map is made from a cube with channel width of 5 km s$^{-1}$ with a threshold of 2 RMS (1.2 mJy beam$^{-1}$) using the CASA command \texttt{immoments}. 

\subsection{HST Data}

We use Pa$\beta$ and I Band maps from \citet{Whitmore_2014} for comparison with the radio continuum. Both images have resolution of $\sim$ 0.2". The I Band image has a pixel size of 0.04" while the Pa$\beta$ image has a larger pixel size of 0.128". In Section \ref{sec:Pbeta_discussion} we use Pa$\beta$ to calculate the total ionized photon number ($Q(H^0)$) and compare it with that derived from 100 GHz continuum.

For Pa$\beta$, we perform continuum subtraction using the HST J Band image. Since the wavelengths of the two filters are close to each other, we adopt a simple model that the intensity at J Band for background stars is the scaled version for that in Pa$\beta$ image. Therefore, the Pa$\beta$ intensity is 
\begin{equation}
\label{eq:contsub}
I_{\mathrm{Pa\beta}} = I_{\mathrm{F128N}} - \beta \cdot I_{\mathrm{F116W}}
\end{equation}
where $I_{\mathrm{F128N}}$ is the Pa$\beta$ intensity before continuum subtraction, $I_{\mathrm{F116W}}$ is the J Band intensity and $I_{\mathrm{Pa\beta}}$ is the intensity after the continuum subtraction. To calculate the $\beta$, we draw apertures around sources that are point like and not associated with any galaxy structures, which are likely background stars. We then measure the fluxes of these sources from F128N and F116W filters. We also measure the background fluxes by drawing apertures close to these background stars. We then plot the background-subtracted fluxes from the two filters and fit a proportional relation to get $\beta$. We calculate $\beta = 0.03$, which is close to $\beta$ value in \citet{Kessler_2020}. We then reproject the J Band image to the Pa$\beta$ and apply Eq. \ref{eq:contsub} to perform the continuum subtraction. 

\begin{figure*}
	\gridline{
		\fig{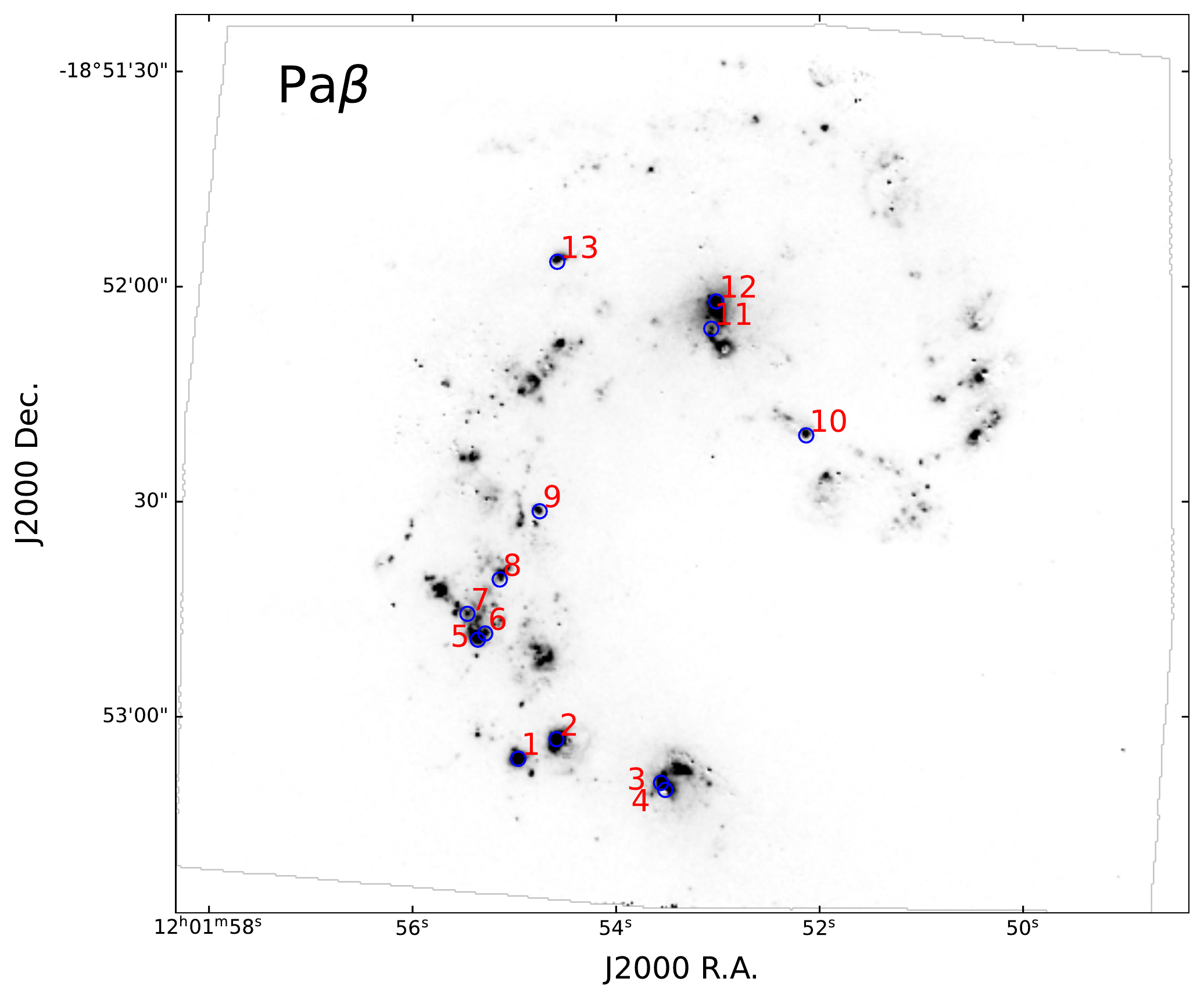}{0.6\linewidth}{}
		\fig{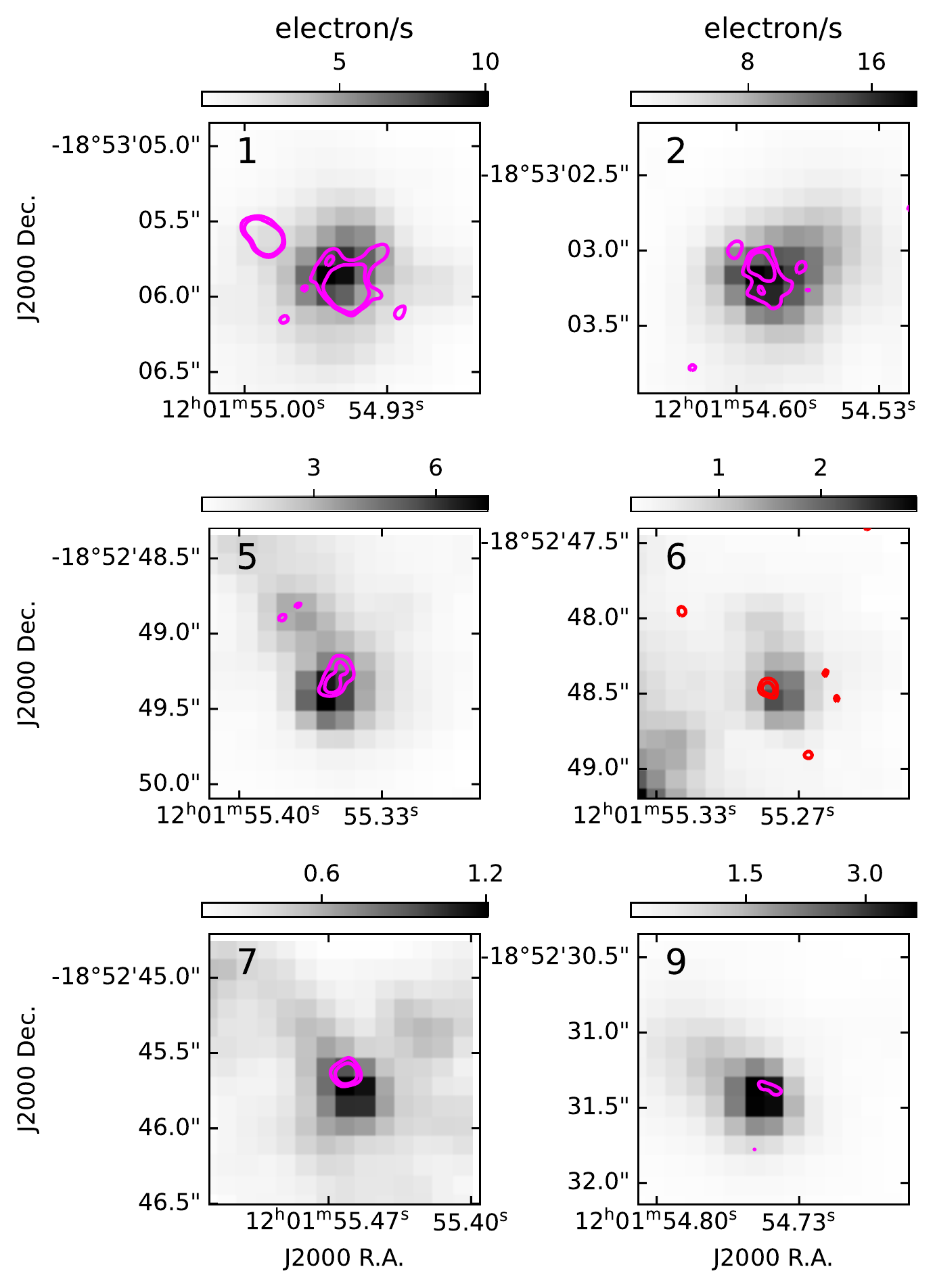}{0.4\linewidth}{}
	}
	\gridline{
	\fig{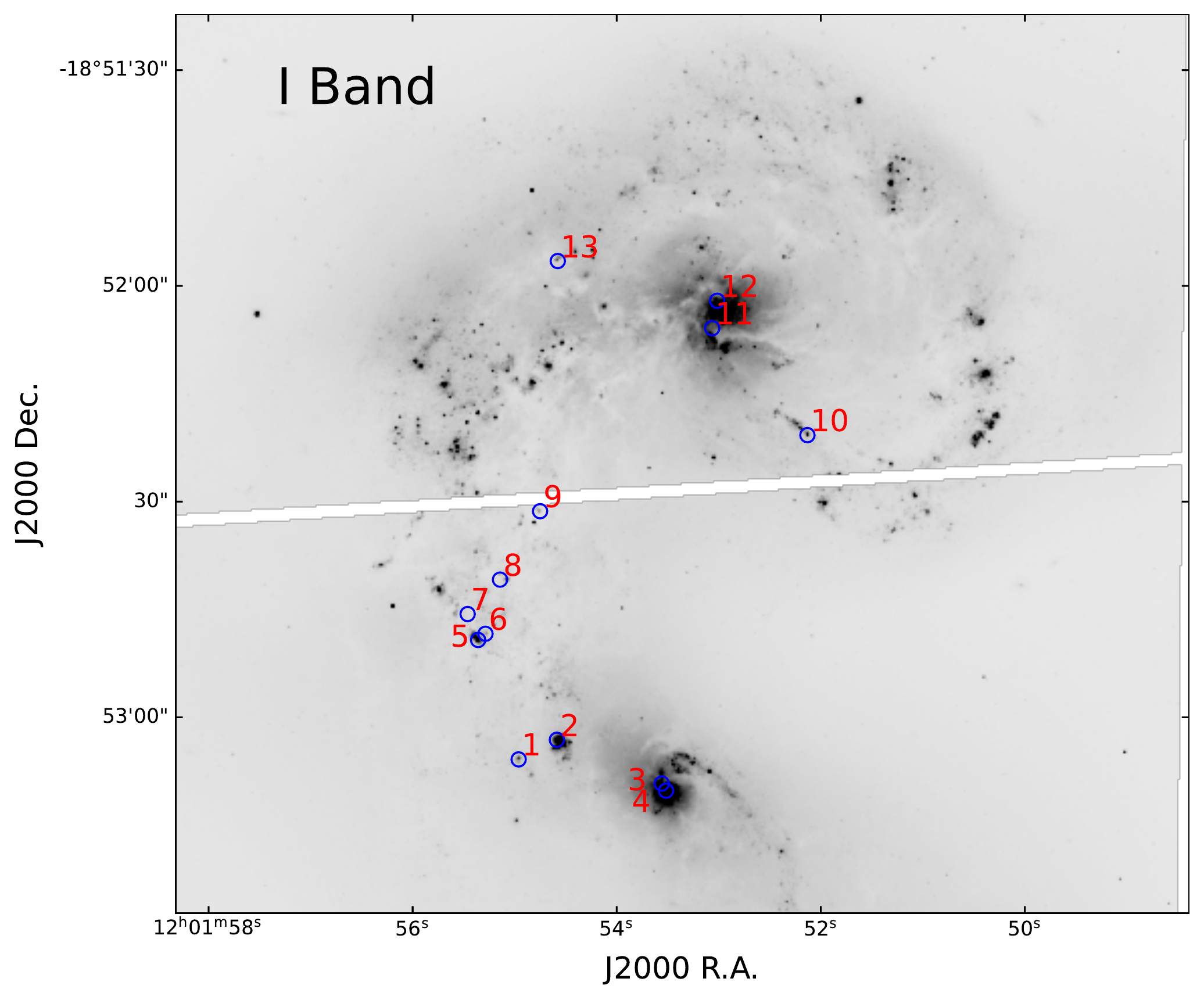}{0.6\linewidth}{}
	\fig{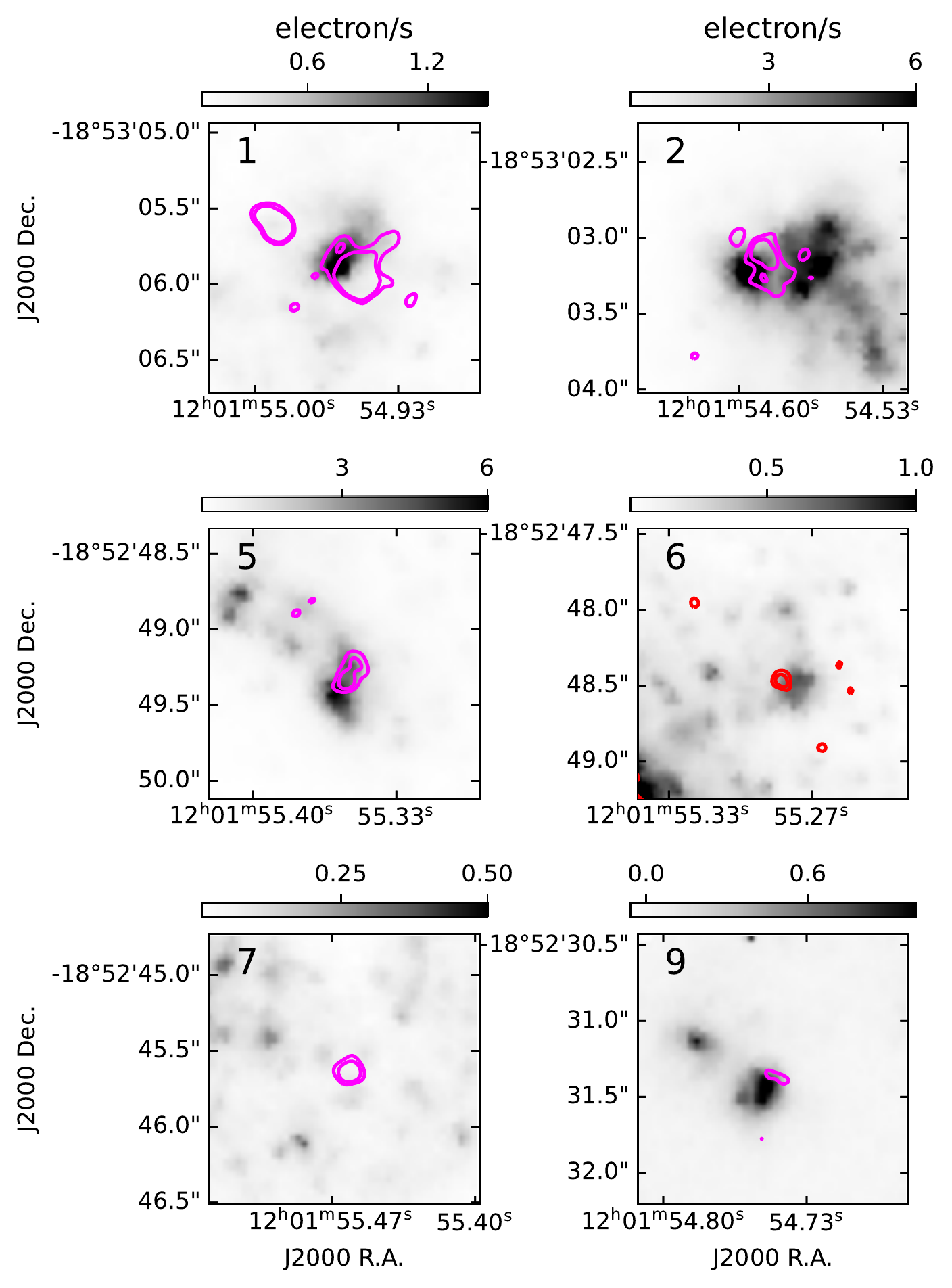}{0.4\linewidth}{}
	}
	\caption{The Paschen $\beta$ (upper) and I band (lower) maps of the Antennae. (Upper Left) The Paschen $\beta$ map of the Antennae \citep{Whitmore_2014} with the positions of the radio continuum sources labeled as blue circles. (Upper Right) Zoom-in Pa$\beta$ images for the individual YMCs. The magenta and red contours are 100 and 345 GHz continuum at 12 pc resolution. The typical offset between the Pa$\beta$ and radio continuum sources is $\sim$ 11 pc. (Lower Left) The HST I Band image for the entire field of view of the Antennae. (Lower Right) Zoom-in I Band image for each YMC.}
	\label{fig:Pbeta_map}
\end{figure*}

\section{Measurements and Derived Quantities} \label{sec:measurements}

\begin{deluxetable*}{clccccc}
	%	\tablenum{1}
	\tablecaption{Measured Quantities of YMCs in the Antennae \label{tab:YMCs_measured}}
	\tablewidth{0pt}
	\tablehead{
		\colhead{Index} & \colhead{Coordinates (J2000)} & \colhead{$S_{\mathrm{100GHz}}$ (mJy)}& \colhead{$S_{\mathrm{340 GHz, dust}}$ (mJy)}&
		\colhead{$\sigma_v$ (km/s)} & \colhead{$T_{\mathrm{kin}}$ (K)} & \colhead{$d_{\mathrm{FWHM}}$} (arcsec)}
	\decimalcolnumbers
	\startdata
	1a & 12h01m54.95s, -18d53m05.98s & 1.2 $\pm$ 0.04  & $<$ 0.04        & $<$ 26 $\pm$ 1 & 40 $\pm$ 2 & 0.085 $\pm$ 0.01 \\
	1b & 12h01m54.99s, -18d53m05.62s & 0.69 $\pm$ 0.04 & 0.89 $\pm$ 0.1  & $<$ 19 $\pm$ 1 & 53 $\pm$ 2 & 0.12 $\pm$ 0.02  \\
	2  & 12h01m54.59s, -18d53m03.10s & 0.41 $\pm$ 0.05 & 0.33 $\pm$ 0.1  & $<$ 17 $\pm$ 2 & 42 $\pm$ 2 & 0.17 $\pm$ 0.04  \\
	3  & 12h01m53.55s, -18d53m09.23s & 0.28 $\pm$ 0.07 & --              & --         & --         & 0.12 $\pm$ 0.03  \\
	4  & 12h01m53.51s, -18d53m10.26s & 0.19 $\pm$ 0.07 & --              & --         & --         & 0.14 $\pm$ 0.05  \\
	5  & 12h01m53.51s, -18d53m10.26s & 0.37 $\pm$ 0.07 & 0.67 $\pm$ 0.1  & $<$ 19 $\pm$ 1 & 48 $\pm$ 2 & 0.28 $\pm$ 0.06  \\
	6  & 12h01m55.28s, -18d52m48.46s & $<$ 0.081       & 0.32 $\pm$ 0.1  & $<$ 23 $\pm$ 2 & 48 $\pm$ 2 & 0.14 $\pm$ 0.03  \\
	7  & 12h01m55.46s, -18d52m45.65s & 0.23 $\pm$ 0.03 & 0.25 $\pm$ 0.07 & $<$ 15 $\pm$ 1 & 38 $\pm$ 2 & 0.06 $\pm$ 0.02  \\
	9  & 12h01m54.75s, -18d52m31.37s & 0.14 $\pm$ 0.06 & 0.19 $\pm$ 0.2  & --         & --         & 0.22 $\pm$ 0.09 
	\enddata
	\tablecomments{(1) Source ID (2) Coordinates. (3) Fluxes at 100 GHz. (4) Estimated dust fluxes at 345 GHz. (5) \cotwo velocity dispersion. All values are upper limits (see Section \ref{sec:linewidth}.)  (6) Kinetic temperature from LTE analysis. (7) Deconvolved FWHM of the source from CASA task \texttt{imfit}. At the distance of the Antennae (22 Mpc), 0.1" = 10.67 pc.}
\end{deluxetable*}

\subsection{Source Identification}

The continuum images from the two ALMA continuum projects are shown in Fig. \ref{fig:cont_image}. As we can see, the continuum images at GMC resolution have higher sensitivity. Therefore, we use the 100 GHz GMC-resolution map, which has the highest sensitivity among all the continuum data, as a guide to find sources in the higher resolution maps by eye. The identified sources are labeled in Fig. \ref{fig:cont_image}. As we can see, there are $\sim$ 10 continuum sources that are likely to be YMCs. The reason we do not observe as many radio continuum sources as optical clusters is mainly due to the limited sensitivity of the radio data. As mentioned in Section \ref{sec:method}, a continuum point source with S/N of 5 corresponds to a 7.2 $\times$ 10$^5$ \solarmass, which is greater than the masses of most star clusters in spiral galaxies. On the other hand, the lifetime of the radio clusters are also much lower than that of optical clusters (see Section \ref{sec:ages} for detailed discussion), which also contributes to the limited number of sources we detected.  

We then match the 100 GHz low-resolution image with the high-resolution image to check if the continuum source is still point-like in the high-resolution image. Source 8 is not detected in the high-resolution image and may be dominated by GMC-scale emission that is filtered out in the high-resolution image. Another interesting source is source 1, which is the brightest source in the continuum map. It separates into two sources in the higher resolution image. Furthermore, source 1b further divides into two sub-sources in our highest resolution image, as shown in Fig. \ref{fig:cluster_summary}. We can see source 1a and 1b have similar velocities (Fig. 2). This comparison suggests these two clusters are close to each other in 3D space and probably interacting with each other within a single GMC. Multiple YMCs within a single GMC have also been observed in NGC 253 \citep{Leroy_2015, Leroy_2018}. Statistical studies \citep[e.g.][]{Grasha_2018, Tsuge_2019} also show that a significant fraction of GMCs will form more than one YMCs. 

We further check the 220 GHz continuum image at GMC scale and the 345 GHz continuum image at star cluster scale. An interesting source is source 6, which is not seen in the 100 GHz continuum image but is seen in both 220 and 345 GHz continuum. Since 100 GHz continuum traces free-free emission from extremely young stars, it seems likely that this source is dominated by a clump of gas with stars yet to form. 

Sources 1a, 1b, 2, 5, 6 and 7 have all been covered by high-resolution observations of 100 and 345 GHz continuum and the \cotwo line with high S/N detections. Therefore, we will focus on these sources in the rest of this paper. Images for these sources are shown in Fig. \ref{fig:cluster_summary}, while the remaining sources are shown in Fig. \ref{fig:cluster_summary_appendix}. The measured properties of all sources are listed in Tables \ref{tab:YMCs_measured} and \ref{tab:YMCs_derived}. The comparison between radio continuum and Pa$\beta$ is shown in Fig. \ref{fig:Pbeta_map}. 

\subsection{Size, Flux and Line Width Measurements}
\label{sec:linewidth}

We determine the sizes of the YMCs using the CASA task \texttt{imfit} on the high-resolution 100 GHz data. This task fits a 2D Gaussian function for a selected encircled region. First, we draw an elliptical aperture around each identified source by eye. We then run the task \texttt{imfit} to get the major axis, minor axis and position angle of the fitted beam. We confirm the fitting results by comparing the half-maximum contour of the source with the fitted apertures on the map. The \texttt{imfit} command also gives us the Gaussian size deconvolved from the beam, which tells us the true source size. We list the major axis of the deconvolved Gaussian in Table \ref{tab:YMCs_measured}. For source 6, since we do not have a detection in 100 GHz continuum, we use the 345 GHz continuum image to fit the source. Note that the beam size of 12 pc is larger than most YMCs in the literature, which have measured sizes of about 2 pc \citep{Leroy_2018}. If the YMCs in the Antennae were to have a similar size, we would expect our derived sizes to be upper limits to the true source size.

%For the star cluster data, we did the fitting on both the robust 0.5 and 2.0 images. We actually use the image with robust value of 0.5, which has the highest resolution of 0.09 arcsec, to determine the size of the continuum source. Many elliptical sources in the robust 2 image break into multiple sources in the robust 0.5 image, such as source 1b, 2, 4, 5 and 9.   

We use an aperture with diameter equal to two times the fitted Gaussian FWHM to measure the flux. The flux uncertainty is calculated as 
$$ Err = \mathrm{RMS} \times \sqrt{N_{\mathrm{beam}}} / \mathrm{pbcor}$$
where RMS is the image noise, $N_{\mathrm{beam}}$ is the number beams in the aperture and $\mathrm{pbcor}$ is the value for the  primary beam response. 

We also measured the line width for the star clusters using the \cotwo map from \citet{Finn_2019} and the same apertures used for the flux measurements. We export the spectrum for each source and then fit a Gaussian to get the velocity dispersion. Many spectra show an extra bump besides the major peak, which comes from gas outside the star cluster. Therefore, we manually set the upper and lower limit for the fitting range to only fit the Gaussian to the major peak. The fits are shown in Fig. \ref{fig:cluster_summary}. In addition, we apply the same method to draw the aperture on sources in the 100 GHz GMC map (see Section \ref{sec:GMC_flux} ). We also overlay the GMC-scale \cotwo spectrum measured in those apertures in the right column of Fig. \ref{fig:cluster_summary}. As we can see, the velocity dispersions at YMC scale are almost the same as those at GMC scale. We suspect this is because the velocity dispersion as measured with \cotwo at YMC scale still traces the overall cloud motions. In this case, the measured velocity dispersion for YMC should be treated as upper limit.

\subsection{Separating Dust Emission and Free-Free Emission} \label{subsec:tables}
At 345 GHz, dust emission is usually expected to be dominant. However, there is still a significant fraction of free-free emission at this frequency for these YMCs. To calculate the dust mass, we need to separate the dust emission from the free-free emission. To begin with, we assume free-free emission dominates the total emission at 100 GHz. This assumption has been shown to hold for YMCs in the Henize 2-10 dwarf galaxy with similar resolution of $\sim$10 pc \citep{Costa_2021}. Free-free emission scales with frequency as a power-law function with index of -0.1 \citep{Ginsburg_2016}. Therefore, we can predict the free-free flux at 340 GHz using
\begin{equation}
S_{\mathrm{345 GHz, ff}} = S_{\mathrm{100GHz}} \left(\frac{345\  \mathrm{GHz}}{100\ \mathrm{GHz}}\right)^{-0.1}
\end{equation} 
where $S_{\mathrm{100GHz}}$ is the continuum flux at 100 GHz. Then the dust flux is just 
\begin{equation}
S_{\mathrm{345 GHz, dust}} = S_{\mathrm{345GHz}} - S_{\mathrm{345 GHz, ff}} 
\end{equation} 
We will use the dust-only flux to calculate the dust and gas mass in the YMCs. 

\subsection{Gas Temperature}

To calculate the dust mass from the dust flux, we need to assume a dust temperature. We assume the dust temperature is equal to the gas kinetic temperature. The gas temperature can then be constrained through the \cotwo observations \citep{Finn_2019} by assuming local thermal equilibrium (LTE). The basic formula to connect the peak brightness temperature and the gas temperature is 
\begin{equation}
T_b = \left[\frac{h\nu /k}{\exp(h\nu/kT_{\mathrm{ex}})-1}- \frac{h\nu /k}{\exp(h\nu/kT_{\mathrm{bg}})-1}\right] \left(1-e^{-\tau}\right)
\end{equation}
where $T_b$ is the peak brightness temperature, $T_{\mathrm{ex}}$ is the excitation temperature, $T_{\mathrm{bg}} = 2.73 K$ is the background temperature, $\nu$ is the observed frequency of the line, $\tau$ is the optical depth, $h$ is Planck's constant and $k$ is the Boltzmann constant. For \cotwo, $h\nu /k = 11.07$ K. We also assume $\tau \rightarrow \infty$. Applying all these assumptions, we can express $T_{\mathrm{kin}}$ as
\begin{equation} \label{eq:LTE_Tkin}
T_{\mathrm{kin}}= T_{\mathrm{ex}} = \frac{11.07}{\ln\left[1+\frac{11.07}{T_b+0.195}\right]}
\end{equation}

The gas temperatures are shown in Table \ref{tab:YMCs_measured}. We see that almost all the sources have $T_{\mathrm{kin}} \sim 40$ K. \citet{Rico-Villas_2020} show that YMCs in NGC 253 have temperatures of 150 - 300 K based on line ratios of HC$_3$N. We note that our physical beam size (12 pc) is much larger than those for NGC 253 (1.9 pc). As discussed in Section 4.1, we probably overestimate the sizes of these YMCs. Therefore, our apertures probably include a large fraction of surrounding gas with lower temperatures. 
 
\subsection{Gas Mass}

The gas mass is calculated from the dust emission after correcting for free-free contamination. We calculate the dust mass based on the equation in \citet{Wilson_2008}, 
\begin{equation} \label{eq: dust_mass}
M_{\text {dust }}=74,220 S_{880} D^{2} \frac{\left(e^{17 / T}-1\right)}{\kappa}\left(M_{\odot}\right)
\end{equation}
where $S_{880}$ is the flux from dust emission at 880 $\mu$m (345 GHz), $D$ is the distance in Mpc, $T$ is the dust temperature in Kelvin and $\kappa$ is the dust emissivity in $\mathrm{cm^{2}\ g^{-1}}$. In this case we assume $\kappa = 0.9\ \mathrm{cm^{2}\ g^{-1}}$ \citep{Wilson_2008}.  The dust mass is highly dependent on the dust temperature. We assume the dust temperature is equal to the gas kinetic temperature (see Section 3.4)

From the dust mass, we then calculate the molecular gas mass based on the gas-to-dust mass ratio: 
\begin{equation} \label{eq: gas_mass}
M_{\mathrm{gas}} = \mathrm{\frac{[Gas]}{[Dust]}} \times M_{\mathrm{dust}}
\end{equation}
We adopt a gas-to-dust mass ratio of 120 from \citet{Wilson_2008}. The gas masses are given in Table \ref{tab:YMCs_derived}. 

\subsection{Extinction}
\label{sec:extinction}

From the gas mass we calculated, we can then derive the gas surface density which is directly related to the optical extinction $A_V$. We can compare $A_V$ from the dust emission with that from optical data to see if they agree with each other. We adapt the equation in \citet{Draine_2003} to calculate the visual extinction $A_V$ from the gas surface density, 
\begin{equation}
A_V = 0.0491 \Sigma_{\mathrm{gas, YMC}}
\end{equation} 
where $\Sigma_{\mathrm{gas, YMC}}$ is the YMC gas surface density derived from the dust continuum flux and deconvolved radius in \coldenunit. The $A_V$ values (Table \ref{tab:YMCs_derived}) show that these sources generally have visual extinctions of hundreds of magnitudes. These extinctions are much larger than the $A_V$ values derived for the optical counterparts of these YMCs (Table \ref{tab:YMCs_optical}). We will discuss this discrepancy in Section \ref{sec:Pbeta_discussion}. 

\subsection{Total Ionizing Photon Number} 
\label{sec:QH0}

We can use both the 100 GHz continuum and the Pa$\beta$ line to calculate the total number of ionizing photons, $Q(H^0)$, since they both trace emission from the ionized gas. We compare the $Q(H^0)$ from two different data sets in Section \ref{sec:Pbeta_discussion}. $Q(H^0)$ can also be used to calculate the stellar mass of young star clusters based on a few assumptions. 

For 100 GHz continuum, we assume the emission is dominated by free-free emission. Therefore, we can calculate the total number of ionizing photons using the equation from \citet{Murphy_2011},
\begin{minipage}{0.45\textwidth}
	\begin{eqnarray}
	\label{eq:Q0_radio}
	\left[\frac{Q\left(H^{0}\right)}{\mathrm{s}^{-1}}\right]=& 6.3 \times 10^{25}\left(\frac{T_{\mathrm{e}}}{10^{4} \mathrm{~K}}\right)^{-0.45}\left(\frac{v}{\mathrm{GHz}}\right)^{0.1} \nonumber \\
	& \times\left(\frac{L_{v}^{\mathrm{T}}}{\mathrm{erg} \mathrm{s}^{-1} \mathrm{~Hz}^{-1}}\right)
	\end{eqnarray}
\end{minipage}
where $Q\left(H^{0}\right)$ is the total number of ionizing photons, $T_e$ is the temperature of the HII region (generally $10^4$ K), $\nu = 100$ GHz is the observed frequency and $L_{v}^{\mathrm{T}}$ is the luminosity of the free-free emission at the observed frequency. 

For Pa$\beta$ data, we use the equation to calculate $Q(H^0)$ from H$\alpha$ \citep{Murphy_2011}. Assume  H$\alpha$/Pa$\beta$ = 17.6 \citep[Case B recombination, $T$=10$^4$ K, and $n_e = 10^4$ cm$^{-3}$; given by ][]{Osterbrock_1989, Cleri_2021} and the equation to calculate $Q(H^0)$ from Pa$\beta$ is given by 
\begin{equation}
\label{eq:Pbeta}
Q_{\mathrm{Pa}\beta}(H^0) = 1.3 \times 10^{13} \  10^{0.4 A_{\mathrm{Pa}\beta}} L_{\mathrm{Pa}\beta}
\end{equation}
where $L_{\mathrm{Pa}\beta}$ is the luminosity of Pa$\beta$ in erg s$^{-1}$ and $A_{\mathrm{Pa}\beta}$ is the extinction for the Pa$\beta$ line in magnitude. $A_{\mathrm{Pa}\beta}$ can be derived from $A_V$ based on the extinction curve in \citet{Calzetti_2000} as 
\begin{equation}
A_{\mathrm{Pa\beta}} = A_V \times \frac{k_{\mathrm{P\beta}}}{k_V} = 0.43 A_V
\end{equation}
where $k_{\mathrm{P\beta}}$ and $k_V$ are the values of reddening curve at the two wavelengths \citep[Eq. 4]{Calzetti_2000}.

\subsection{Stellar Mass}
From $Q\left(H^{0}\right)$, we can calculate the stellar mass using the equation from \citet{Leroy_2018}, 
\begin{equation}
M_{\star} = \frac{Q\left(H^{0}\right)}{4 \times 10^{46}}\  M_{\odot} 
\end{equation}
This equation assumes a Kroupa initial mass function with maximum stellar mass of 100 $M_{\odot}$ and also assumes YMCs have a single stellar population (SSP). The stellar masses calculated from the 100 GHz fluxes in Table 2 are given in Table 3. 

\section{Star cluster properties}

\begin{deluxetable*}{ccccccc}
%	\tablenum{2}
	\tablecaption{Derived Physical Properties of YMCs \label{tab:YMCs_derived}}
	\tablewidth{0pt}
	\tablehead{
		\colhead{Index} & \colhead{$\log_{10} M_{\star}$ } &  \colhead{$\log_{10} M_{\text{gas}}$ }&
		\colhead{$\log_{10} M_{\text{vir}}$ } & \colhead{$\log_{10} \Sigma_{\mathrm{tot}}$} & \colhead{$R_h$ } & \colhead{$A_V$ }  \\
		\nocolhead{Common} & 	\colhead{(\solarmass)} & \colhead{(\solarmass)} & \colhead{(\solarmass)} & \colhead{(\solarmass pc$^{-2}$)} & \colhead{(pc)} & \colhead{(10$^2$ mag)} }
	\decimalcolnumbers
	\startdata
	1a & 6.2 $\pm$ 0.01 & $<$ 5.4        & $<$ 6.8 $\pm$ 0.06 & 4.4 $\pm$ 0.2  & 4.5 $\pm$ 0.5 & $< 2$\\
	1b & 6.0 $\pm$ 0.03 & 6.8 $\pm$ 0.05 & $<$ 6.7 $\pm$ 0.09 & 4.7 $\pm$ 0.2 & 6.4 $\pm$ 1.1 & 22 $\pm$ 8 \\
	2  & 5.8 $\pm$ 0.05 & 6.4 $\pm$ 0.15 & $<$ 6.8 $\pm$ 0.13 & 4.1 $\pm$ 0.2  & 9.1 $\pm$ 2.1 & 4.5 $\pm$ 2.6\\
	5  & 5.7 $\pm$ 0.08 & 6.7 $\pm$ 0.08 & $<$ 7.0 $\pm$ 0.1  & 3.9 $\pm$ 0.2 & 15 $\pm$ 3   & 3.2 $\pm$ 1.5\\
	6  & $<$ 5.1        & 6.3 $\pm$ 0.15 & $<$ 6.9 $\pm$ 0.1  & 4.1 $\pm$ 0.2 & 7.5 $\pm$ 1.6 & 6.1 $\pm$ 3.3\\
	7  & 5.5 $\pm$ 0.05 & 6.3 $\pm$ 0.13 & $<$ 6.3 $\pm$ 0.2  & 4.7 $\pm$ 0.3  & 3.2 $\pm$ 1.1 & 28 $\pm$ 20 
	\enddata
	\tablecomments{(1) Source ID. (2) Stellar mass. (3) Dust mass. (3) Gas mass derived from dust continuum (see Section 3.5). (4) Virial mass as upper limit. (5) Total surface density (gas+star) of YMCs (6) Half-light radius (7) Optical extinction at \textit{V} band}
\end{deluxetable*}

\subsection{Ages of the YMCs}
\label{sec:ages}

We can estimate the ages of the YMCs by comparing the number of clusters at optical wavelengths with the number observed in the radio. This method has been used previously for estimating ages of YMCs in the dwarf galaxy Henize 2-10 \citep{Kobulnicky_1999, Johnson_2003}.
The number ratio is roughly the age ratio of the two populations if we assume a constant SFR during this time period. 
\citet{Zhang_2001} estimate there are about 1600 clusters with ages smaller than 16 Myr and masses greater than $10^4$ \solarmass. For our 100 GHz GMC map, the S/N=5 cutoff is $\sim 7.2 \times 10^4$ \solarmass. To estimate the number of optical clusters with masses greater than that value, we assume the cluster mass function has a slope of -2 \citep{Krumholz_2019}. In this case, we expect about 200 optical clusters with masses greater than $7.2 \times 10^4$ \solarmass and ages less than 16 Myr. Our actual 100 GHz GMC map reveal 17 continuum sources greater than that value. Based on that number, we would expect the age of these radio YMCs to be $\sim$ 1 Myr. Various HST studies of YMCs have adopted similar statistical counting methods and find the time for star clusters to dissociate from their host GMCs is about 2--3 Myr \citep{Hollyhead_2015, Grasha_2018, Hannon_2019}. If this timescale is also true for the Antennae, we would expect feedback has not been effective for most of our radio sources. Therefore, they should still be forming stars.  

\subsection{Size-Mass relation}

\begin{figure}
	\plotone{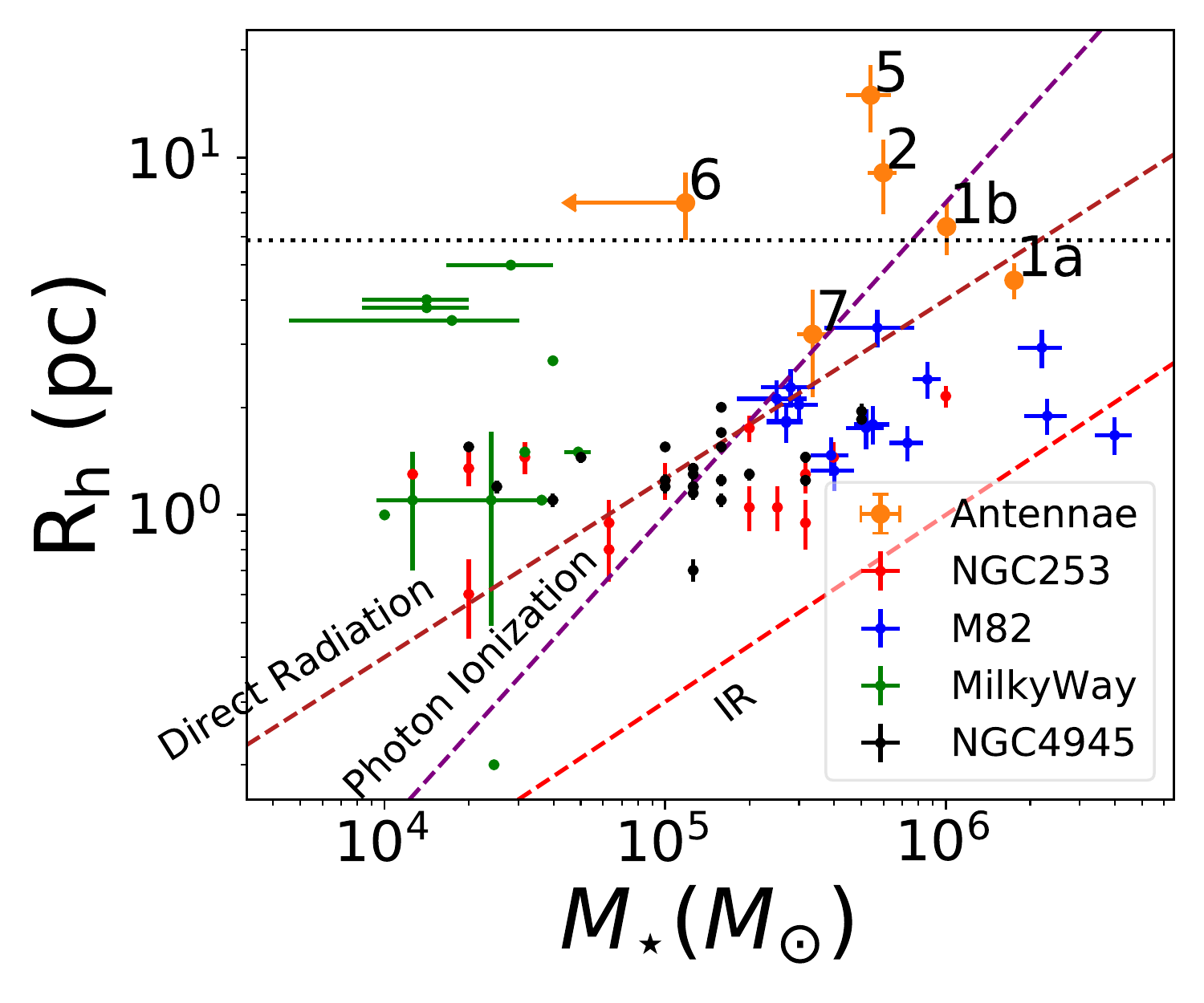}
	\caption{The half-light radius versus stellar mass for YMCs in the Antennae, NGC 253 \citep{Leroy_2018}, M82 \citep{McCrady_2007}, Milky Way \citep{Krumholz_2019} and NGC 4945 \citep{Emig_2020}. The horizontal dotted line marks our YMC resolution (0.11"). The diagonal dashed lines specify the area where the feedback from direct radiation (brown), photoionization (purple) and IR radiation (red) is effective for dispersing the surrounding gas \citep{Krumholz_2019}. For direct radiation and photoionization, the effective area is leftward and for IR radiation, the effective area is rightward. }
	\label{fig:size_mass}
\end{figure}

We plot our measurements of the YMCs in the size-mass relation along with literature data from other galaxies (Fig \ref{fig:size_mass}).  According to \citet{Leroy_2018} and \citet{Levy_2020}, very young YMCs are generally very compact with radii of 1 $\sim$ 2 pc. However, from the fit to the size-mass relation in the LEGUS sample \citep{Brown_2021}, we would expect a radius of $\sim$ 5 pc for a YMC with stellar mass of $10^6$ \solarmass. This radius is consistent with what we measured for some of the YMCs in the Antennae. Note that the \citet{Brown_2021} relation for young star clusters does not extend to 10$^6$ \solarmass. 

We also plot dashed lines from \citet{Krumholz_2019} to show regions where feedback is effective. As we can see, most of the YMCs lie in the area where direct radiation or photoionization feedback is effective. This suggests two possibilities. One is that feedback is actually effective for these YMCs. As shown in Table \ref{tab:YMCs_derived}, most YMCs except for source 1a have gas fractions greater than 50\%. According to various observations \citep{Whitmore_2002, Whitmore_2014, Hannon_2019, Chevance_2020}, the timescale for feedback to disperse the gas is around 1 -- 3 Myr. As we have calculated, these YMCs should generally have ages of $\sim$ 1 Myr. Therefore, although feedback is expected to play a role, it has not cleared all the gas surrounding the YMCs. 

On the other hand, we might overestimate the sizes of these YMCs. As shown in Fig. \ref{fig:size_mass}, our resolution limit is higher than most of YMCs with similar masses. As an example, suppose the radii are overestimated by a factor of 4: cluster 1a would then fall on the line for IR radiation feedback to be effective, while the rest of the clusters would lie in the region where no feedback is effective. However, we note that the feedback region in the size-mass diagram is one projection of a complex process. In \citet{Levy_2020}, they find outflows in YMCs which they do not expect to be experiencing feedback given their location in the size-mass diagram.

\subsection{Virial Mass}

We use the equation in \citet{Bolatto_2013} to calculate the virial mass,
\begin{equation} \label{eq:Leroy_vir}
M_{\mathrm{vir}} = 1061 \ R_{h} \ \sigma_v^2
\end{equation}
where $M_{\mathrm{vir}}$ is the virial mass in \solarmass,  $R_{h}$ is the deconvolved half-light radius in pc and $\sigma_v$ is the measured velocity dispersion in km s$^{-1}$ of the source. This equation assumes uniform density.

\begin{figure}
	\plotone{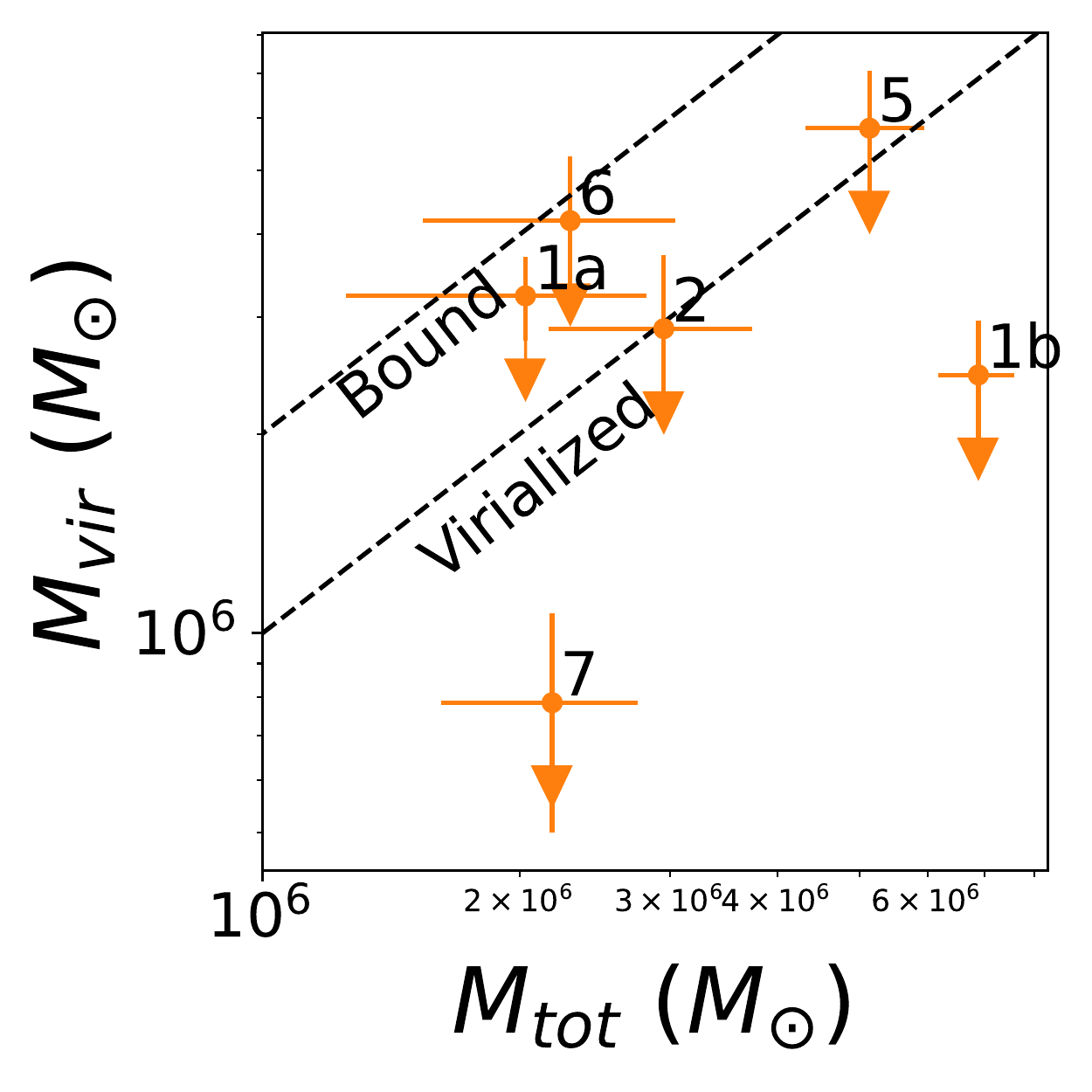}
	\caption{The comparison between the total mass and the virial mass. The two diagonal dashed lines mark the boundaries below which systems are bound or virialized. The offset between the dashed lines is a factor of 2. 
	We can see that most of sources are bound systems. }
	\label{fig:vir}
\end{figure}

Fig. \ref{fig:vir} shows the comparison between virial mass and total mass. We can see that the virial masses for all the YMCs are smaller than $2 \times M_{\mathrm{tot}}$, which implies that these YMCs are likely gravitationally bound systems. We need to note that the virial masses of the YMCs should be treated as upper limits because our measured velocity dispersion is tracing the overall cloud motion instead of the dispersion inside the YMC. This further confirms that all these sources should be gravitationally bound system and may be virialized.  

Although these YMC+gas systems are currently gravitationally bound, 
whether the final star cluster will be gravitationally bound or not 
depends on the fraction of the gas mass that is eventually turned into 
stars. Thus, the boundedness of these very young embedded YMCs is not 
necessarily inconsistent with the results from \citet{Matthews_2018}, 
who found that only a small fraction of the optically visible young star clusters in the Antennae are likely to remain gravitationally bound.

\subsection{Comparison with Optical Data}
\label{sec:Pbeta_discussion}

\begin{deluxetable*}{ccccccccc}
\tablecaption{Corresponding optical YMCs 
\label{tab:YMCs_optical}}
\tablewidth{0pt}
\tablehead{
	\colhead{Index} & \colhead{Region} & \colhead{HST ID} &  \colhead{$\log_{10} M_{\star}$}&
	\colhead{Age} & \colhead{A$_V$} & \colhead{ $S_{\mathrm{Pa{\beta}}}$} & \colhead{$S_{\mathrm{100GHz, smooth}}$}  & \colhead{Cross ID} \\
	\nocolhead{Common} & 	\nocolhead{Common} & \nocolhead{Common} & \colhead{(\solarmass)} & \colhead{(Myr)} & \colhead{(mag)} & \colhead{(10$^{-14}$ erg s$^{-1}$)} & \colhead{(mJy)} & \nocolhead{Common} }
\decimalcolnumbers
\startdata
1a                                              & SGMC 4/5 & 14612  & 6.8                         & 1.0   & 7.3   & 1.92                             & 2.1 $\pm$ 0.05                & B1       \\
2                                               & SGMC 4/5 & 15492  & 6.6                         & 2.5   & 3.0   & 4.1                             & 1.3 $\pm$ 0.07                & B        \\
5                                               & SGMC 1   & 19330  & 5.5                         & 1.0   & 1.0   & 0.84                             & 0.4 $\pm$ 0.04                & D        \\
6                                               & SGMC 1   & 19807  & --                          & 8     & 5.0   & 0.24                             & $<$ 0.075                     & D1       \\
7                                               & SGMC 1   & --     & --                          & 3.5   & 4.1   & 0.16                             & 0.61 $\pm$ 0.05               & D2       \\
9                                               & LT       & 3475   & 6.0                    & 1.0   & 4.2   & 0.36                             & 0.35 $\pm$ 0.07               & E3           
\enddata
\tablecomments{(1) Source ID from Table \ref{tab:YMCs_measured}. (2) Regions defined in \citet{Whitmore_2014} based on \cothree map. (3) HST IDs for star clusters identified in \citet{Whitmore_2010}(source 1a, 2, 5, 6) or \citet{Whitmore_2002}(source 9). (4) Stellar mass of the star clusters from \citet{Whitmore_2010} or \citet{Whitmore_2002}. (5) Age of the star clusters from \citet{Whitmore_2014}. (6) Extinction at $V$ band from \citet{Whitmore_2014}. (7) The flux measured for Pa$\beta$ sources. (8) The flux measured for 100 GHz radio sources smoothed to the resolution of 0.201". (9) Cross IDs of K Band sources identified in  \citet{Gilbert_2007}.  }
\end{deluxetable*}

\begin{figure*}[htb!]
	\gridline{
		\fig{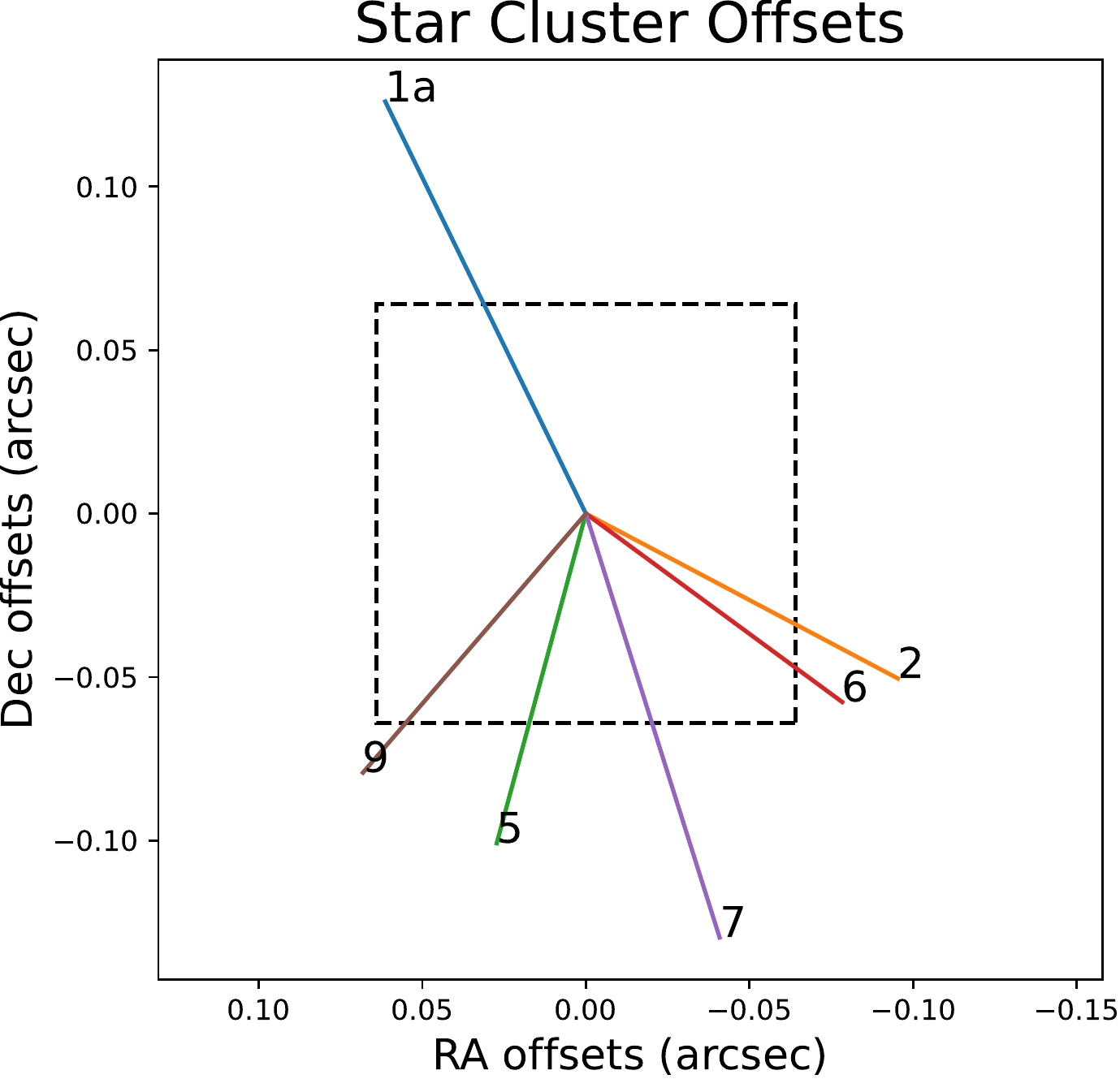}{0.5\textwidth}{}
		\fig{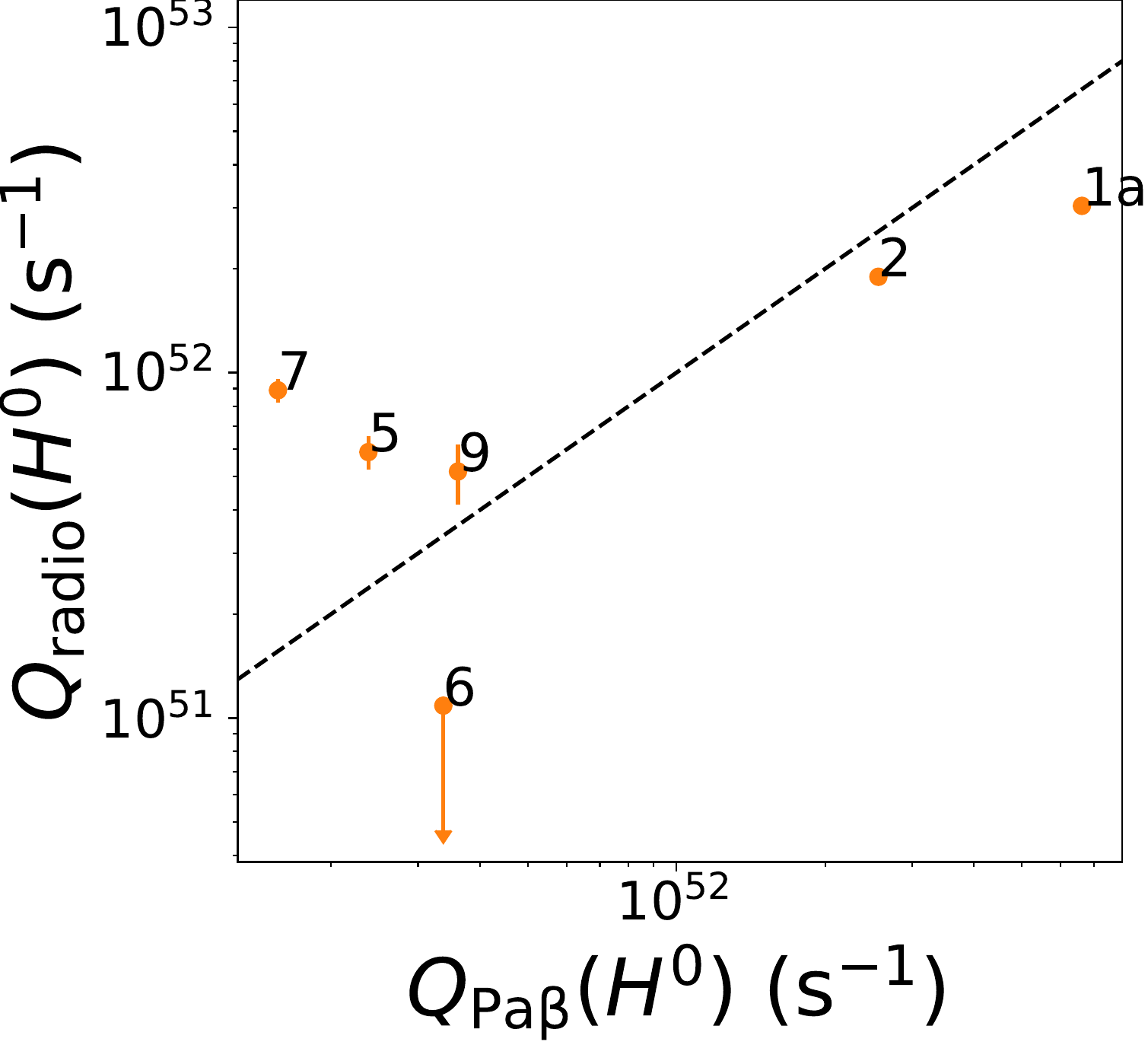}{0.5\textwidth}{}
	}
	\caption{(Left) The coordinate offsets between Pa$\beta$ sources and radio continuum sources. The dashed square shows the pixel size of the Pa$\beta$ image (0.128"). (Right) The comparison of the total ionizing photon numbers derived from Pa$\beta$ and 100 GHz images. The dashed line shows the 1-to-1 relation. }
	\label{fig:Pbeta_radio}
\end{figure*} 

\begin{figure*}[htb!]
	\epsscale{0.7}
	\plotone{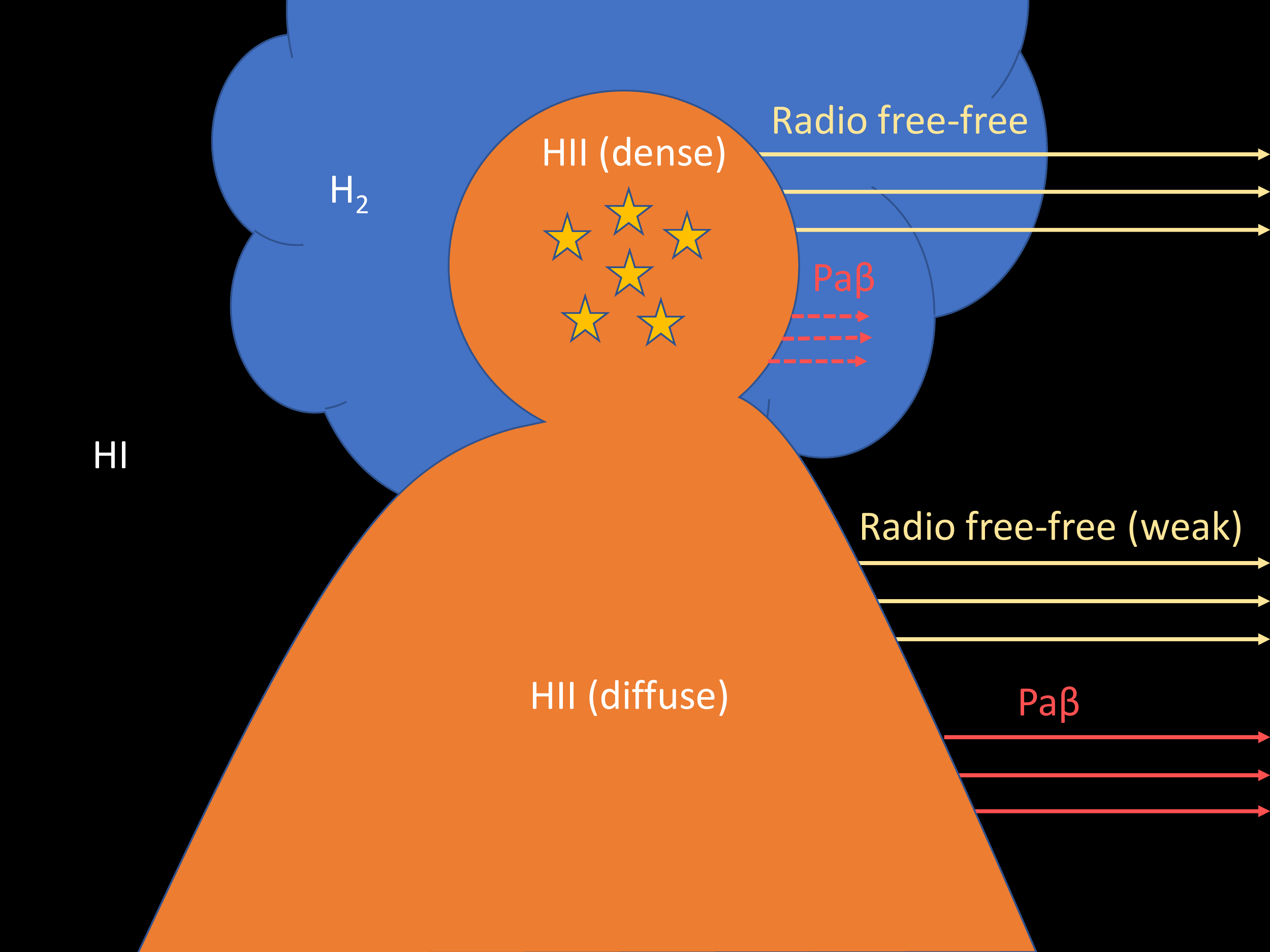}
	\caption{A schematic showing an HII region ionized by a star cluster that is at the edge of the cloud. In addition to the HII region inside the cloud, some photons leak out and ionize some of the HI gas outside of the cloud. The HII region inside the cloud is smaller than the HII region outside the cloud due to the higher gas pressure in the cloud. Both dense and diffuse HII regions produce radio free-free continuum and Pa$\beta$ emission. We cannot see the Pa$\beta$ emission from the HII region inside the cloud due to the high extinction. Outside the cloud, we can see both radio free-free emission and Pa$\beta$ emission. However, the radio free-free emission might not be detectable due to the low density of the HII region.}
	\label{fig:HII_diagram}
\end{figure*}

Based on the multi-wavelength comparison in \citet{Whitmore_2014}, we can see that most of our sources have optical counterparts. The properties of these optical YMCs are summarized in Table \ref{tab:YMCs_optical}. As we mentioned in Section \ref{sec:extinction}, the $A_V$ derived from the optical data is $\sim$ 100 times smaller than that derived from the dust continuum. This difference suggests the optical source and radio source might just happen to be along the same line of sight. Furthermore, the radio continuum data used in \citet{Whitmore_2014} only has a resolution of $\sim$ 0.5". With our new data, we can make a more precise comparison of the coordinates of the optical and radio sources. We use Pa$\beta$ for this comparison since both 100 GHz radio continuum and Pa$\beta$ trace emission from the ionized gas. 

The Pa$\beta$ and I Band maps for the Antennae and the individual sources are shown in Fig. \ref{fig:Pbeta_map}. As we can see, the Pa$\beta$ image looks quite similar to the I band image. The individual I band sources and Pa$\beta$ sources also match and have consistent offsets from the radio sources except for source 7, which does not have I band detection which we hypothesis is due to the extinction. To make a more quantitative comparison between coordinates of radio sources and those of HST sources, we apply \texttt{imfit} on Pa$\beta$ sources to get the central coordinates and compare them with those derived from radio continuum. The offsets between the two central coordinates are shown in the left panel of Fig. \ref{fig:Pbeta_radio}. As we can see, these offsets are not in a consistent direction and so we cannot shift the coordinates of the Pa$\beta$ image to align the peak of the Pa$\beta$ sources with the radio sources. The offsets are generally larger than 0.1", which translates to a physical distance of 11 pc. In comparison, the offset between the 100 GHz and 345 GHz image is typically less than 0.03". 

On the other hand, as mentioned in Section \ref{sec:QH0}, we can use both images to calculate $Q(H^0)$. If the $Q(H^0)$ from the two sets of data agree with each other, then it seems likely that the Pa$\beta$ emission and 100 GHz continuum are from the same physical source instead of from two sources that happen to lie along the same line of sight. We smooth the radio continuum image to a resolution of 0.21" to match the Pa$\beta$ image and then use Eq. \ref{eq:Q0_radio} and \ref{eq:Pbeta} to calculate the total number of ionizing photons from 100 GHz radio continuum and Pa$\beta$. 
As shown in the right panel of Fig. \ref{fig:Pbeta_radio}, the total numbers of ionizing photons derived from the two datasets generally agree with each other within a factor of 2. This agreement suggests the radio continuum and Pa$\beta$ might actually trace the same physical source, despite the offset in their coordinates.  

One possible explanation is that clumpiness or density gradients in the surrounding molecular cloud is allowing photons from the YMC to leak out of the cloud and ionize the surrounding HI gas (Fig. \ref{fig:HII_diagram}). This has been proposed as the 'blister' model \citep{Israel_1978} to explain the spatial and velocity offset between CO clouds and HII regions in Milky Way. In this case, high extinction inside the cloud could prevent us from observing the Pa beta emission produced there, while the relatively low sensitivity of the radio continuum data could prevent us from detecting radio continuum emission from the HII region outside the cloud. It would explain the spatial offset between the peaks at different wavelengths and also why we have different A$_V$ values from optical data and 345 GHz dust emission. 

For source 6, the ionizing photon counts from Pa$\beta$ are much higher than the upper limit derived from the 100 GHz radio continuum image. A possible scenario is that this source has already emerged from the cloud and heats the nearby GMC without ionizing much of the cloud. In this scenario, the detected 345 GHz continuum emission comes primarily from heated dust from the edge of the GMC, while any radio continuum emission from the ionized gas seen in Pa$\beta$ is too faint to be detected. Another interesting object is source 7, which has lower $Q(H^0)$ derived from Pa$\beta$ emission than source 6, but which is detected in 100 GHz continuum. Since this source does not have I band detection, we suspect this source is still quite embedded in the parent cloud and only has small amount of ionizing photons leaked out to generate the Pa$\beta$ emission. 

\subsection{Missing Proto Star Cluster - Firecracker}

\citet{Whitmore_2014} identified a candidate of proto star cluster called the Firecracker in the SGMC 2 region. It is luminous and compact in \cothree\ but without any associate radio counterpart at 3.6 cm \citep{Johnson_2015}, which suggests it is at the very beginning stage of forming stars. Therefore, we do not expect the source to appear in our 100 GHz map. On the other hand, we do expect it to have strong dust emission at higher frequencies. However, as we see in 345 GHz map (Fig \ref{fig:cont_image}), there is no signal detected in SGMC 2 region. 

\citet{Johnson_2015} detected the Firecracker at 345 GHz with a resolution of 0.5" (53 pc). The peak intensity of the dust continuum reported is 9.8$\times 10^{-4}$ Jy beam$^{-1}$. If we assume it is a perfect point source, the same peak intensity at 0.11" resolution would  give us a S/N of 25, which is clearly not the case. On the other hand, if we assume the dust emission is uniformly distributed over the 53 pc area, we would expect a S/N of 1 for the emission peak, which agrees better with what we observe. This analyses clearly suggests that Firecracker has structure on GMC scales. Furthermore, we would expect the dust temperature to be quite cold throughout the whole area due to the lack of stellar radiation. Therefore, the dust would not be as luminous as our YMC candidates and thus the Firecracker would not appear as strong point source. 

\section{Comparison with GMC properties}

\begin{deluxetable*}{clcccccc}
	%	\tablenum{1}
	\tablecaption{Measured properties of selected GMCs in the Antennae \label{tab:GMCs}}
	\tablewidth{0pt}
	\tablehead{
		\colhead{Index} & \colhead{Coordinates} & \colhead{$S_{\mathrm{100GHz}}$}& \colhead{$S_{\mathrm{220 GHz, dust}}$}&
		\colhead{$S_{\mathrm{345 GHz, dust}}$} & \colhead{$T_{\mathrm{kin}}$} &  \colhead{$\log_{10} M_{\text{gas}}$} & \colhead{$\log_{10} \Sigma_{\text{gas}}$} \\
 		\nocolhead{Common} & 	\colhead{(J2000)} & \colhead{(mJy)} & \colhead{(mJy)} & \colhead{(mJy)} & \colhead{(K)} &  (\solarmass) & (\coldenunit)}
	\decimalcolnumbers
	\startdata
1  & 12h01m54.96s, -18d53m05.86s & 3.5 $\pm$ 0.04  & 1.5 $\pm$ 0.1   & 6.9 $\pm$ 0.5 & 21.5 $\pm$ 0.3 & 7.86 $\pm$ 0.03 & 3.90 $\pm$ 0.03 \\
2  & 12h01m54.59s, -18d53m03.10s & 1.2 $\pm$ 0.03  & 0.57 $\pm$ 0.09 & 2.6 $\pm$ 0.4 & 12.3 $\pm$ 0.4 & 7.69 $\pm$ 0.07 & 3.85 $\pm$ 0.07 \\
3  & 12h01m53.55s, -18d53m09.23s & 0.25 $\pm$ 0.03 & 0.42 $\pm$ 0.08 & 1.9 $\pm$ 0.4 & 16.8 $\pm$ 0.4 & 7.40 $\pm$ 0.09 & 3.68 $\pm$ 0.09 \\
4  & 12h01m53.51s, -18d53m10.26s & 0.12 $\pm$ 0.02 & $<$  0.07       & $<$ 0.3       & 13.8 $\pm$ 0.4 & $<$ 6.71        & $<$ 3.17        \\
5  & 12h01m53.51s, -18d53m10.26s & 0.52 $\pm$ 0.03 & 0.57 $\pm$ 0.1  & 2.6 $\pm$ 0.4 & 22.8 $\pm$ 0.3 & 7.42 $\pm$ 0.07 & 3.55 $\pm$ 0.07 \\
6  & 12h01m55.28s, -18d52m48.46s & 0.21 $\pm$ 0.04 & 0.37 $\pm$ 0.1  & 1.7 $\pm$ 0.5 & 15.8 $\pm$ 0.4 & 7.37 $\pm$ 0.13 & 3.45 $\pm$ 0.12 \\
7  & 12h01m55.46s, -18d52m45.65s & 0.51 $\pm$ 0.03 & 0.71 $\pm$ 0.09 & 3.3 $\pm$ 0.4 & 17.6 $\pm$ 0.4 & 7.61 $\pm$ 0.05 & 3.87 $\pm$ 0.05 \\
8  & 12h01m55.14s, -18d52m40.86s & 0.18 $\pm$ 0.03 & 0.45 $\pm$ 0.09 & 2.1 $\pm$ 0.4 & 14.3 $\pm$ 0.4 & 7.51 $\pm$ 0.09 & 3.70 $\pm$ 0.09 \\
9  & 12h01m54.75s, -18d52m31.37s & 0.19 $\pm$ 0.03 & 0.21 $\pm$ 0.08 & 1.0 $\pm$ 0.4 & 9.9 $\pm$ 0.5  & 7.42 $\pm$ 0.16 & 3.83 $\pm$ 0.16 \\
10 & 12h01m52.13s, -18d52m20.76s & 0.19 $\pm$ 0.02 & 0.24 $\pm$ 0.07 & 1.1 $\pm$ 0.3 & 15.2 $\pm$ 0.4 & 7.21 $\pm$ 0.13 & 3.62 $\pm$ 0.13 \\
11 & 12h01m53.06s, -18d52m05.88s & 0.16 $\pm$ 0.03 & 0.54 $\pm$ 0.08 & 2.5 $\pm$ 0.4 & 19.0 $\pm$ 0.3 & 7.46 $\pm$ 0.07 & 3.76 $\pm$ 0.06 \\
12 & 12h01m53.02s, -18d52m02.07s & 0.46 $\pm$ 0.03 & 1.2 $\pm$ 0.1   & 5.7 $\pm$ 0.5 & 27.5 $\pm$ 0.3 & 7.70 $\pm$ 0.04 & 3.81 $\pm$ 0.03 \\
13 & 12h01m54.58s, -18d51m56.55s & 0.22 $\pm$ 0.03 & 0.43 $\pm$ 0.13 & 2.0 $\pm$ 0.6 & 13.1 $\pm$ 0.4 & 7.55 $\pm$ 0.13 & 3.87 $\pm$ 0.13  
	\enddata
	\tablecomments{(1) Source ID. (2) Coordinates. (3) GMC fluxes at 100 GHz. (4) GMC dust fluxes at 220 GHz. (5) GMC dust fluxes at 345 GHz. (6) Peak gas kinetic temperature. (7) GMC gas mass derived from dust continuum. (see Section 5.1) (8) GMC surface density derived from dust continuum. }
\end{deluxetable*}

\subsection{Flux at GMC scales}
\label{sec:GMC_flux}

\begin{figure*}
	\plottwo{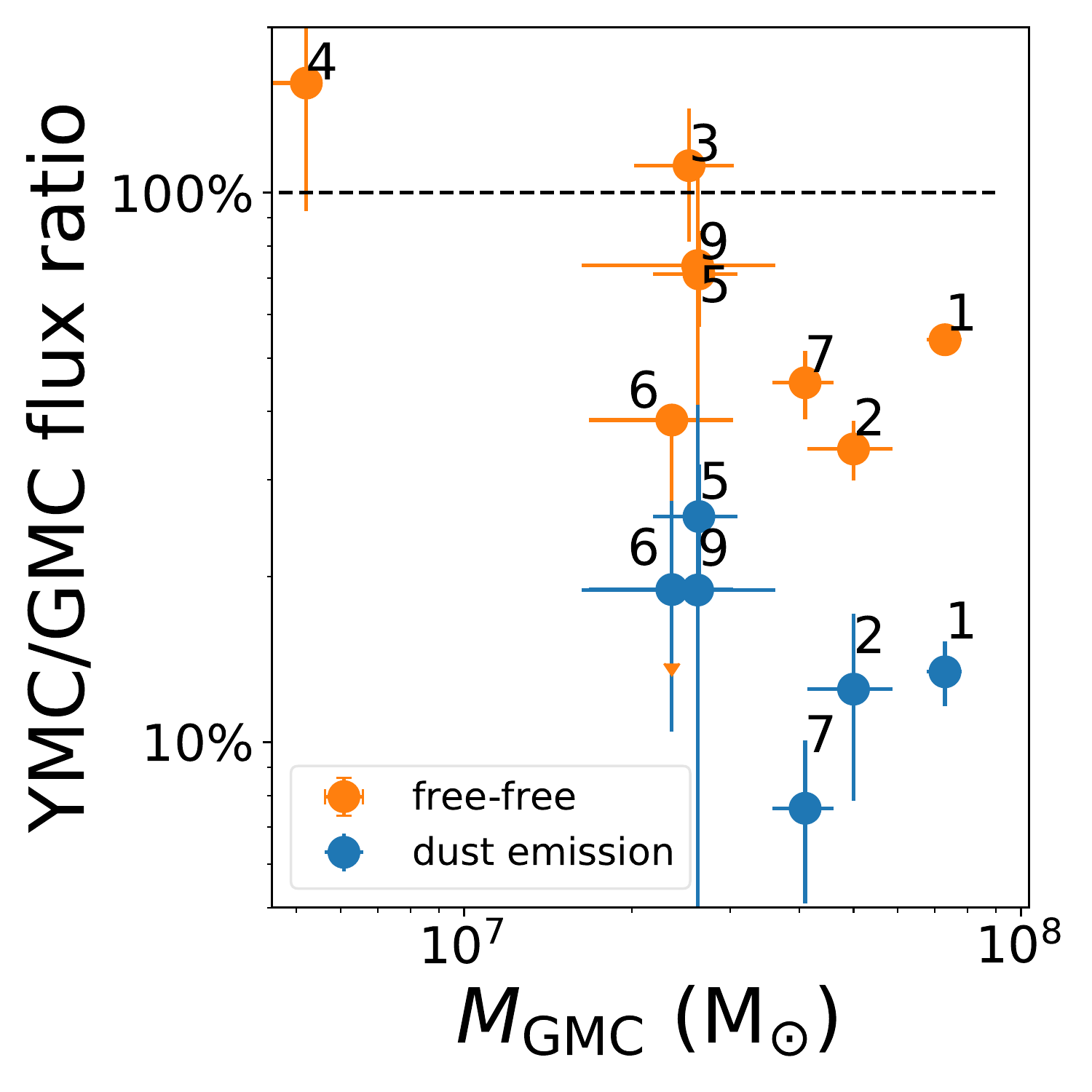}{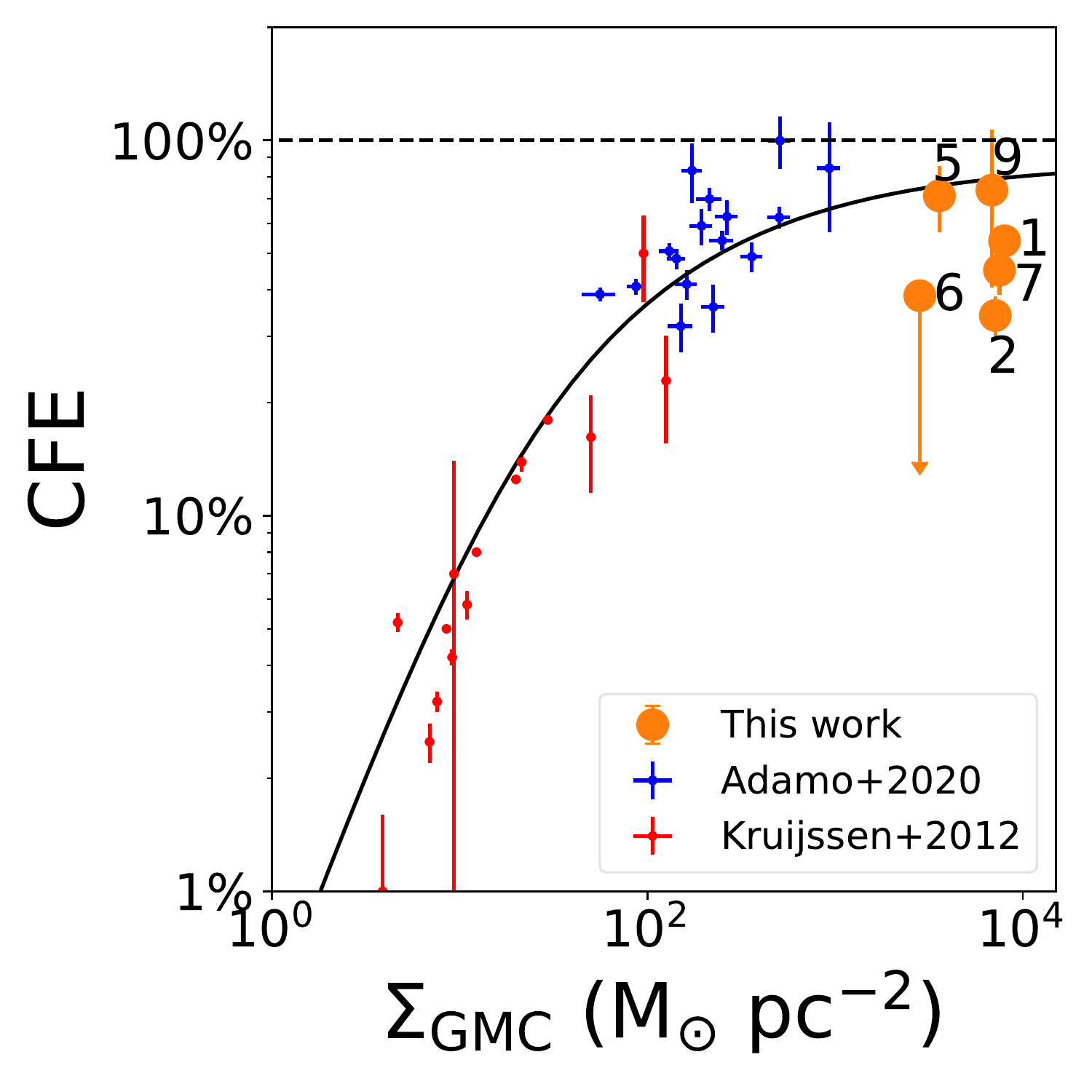}
	\caption{(Left) The ratio of flux at YMC scales to that at GMC scales versus the GMC mass. Flux ratios are shown for both free-free emission and dust emission. (Right) The YMC-to-GMC ratio for free-free emission, which is equivalent to the cluster formation efficiency (CFE) versus the gas surface density. The mean gas surface density for individual YMCs from the Antennae is taken to be the gas surface density at GMC scales (Table \ref{tab:GMCs}). The blue and red points are CFEs for individual galaxies compiled by \citet{Adamo_2020} and \citet{Kruijssen_2012}. The solid curve is the theoretical prediction of CFE from \citet{Kruijssen_2012}. We can see that CFEs for individual GMCs in the Antennae generally agree with the observations and theoretical predictions for entire galaxies. }
	\label{fig:flux_ratio}
\end{figure*}

To measure the fluxes at GMC scales, we apply a similar procedure as described in Section 3.2 with the aperture determined through 2D Gaussian fitting to the 100 GHz GMC-scale continuum image. We use a similar procedure to split the free-free emission and dust emission as described in Section 3.2, although now we are calculating the dust flux at 220 GHz instead of 345 GHz. The free-free and dust emission at 220 GHz are calculated as 

\begin{eqnarray}
S_{\mathrm{220 GHz, ff}} &=& S_{\mathrm{100GHz}} \left(\frac{220\  \mathrm{GHz}}{100\ \mathrm{GHz}}\right)^{-0.1} \\
S_{\mathrm{220 GHz, dust}} &=& S_{\mathrm{220 GHz}} - S_{\mathrm{220 GHz, ff}}
\end{eqnarray} 

Previous studies \citep[e.g.][]{Whitmore_2010} have used continuum data with $\sim$ 60 pc resolution to study star clusters, which generally have diameters of several pc. This approach will potentially include emission from outside the star cluster. A comparison between our 61 pc and 12 pc resolution data will enable us to quantify this bias. In addition, the ratio between fluxes from the two spatial scales will tell us how concentrated the star formation is in each individual GMC. After we split the dust emission from the free-free emission, we can calculate the GMC gas mass based on Eq. \ref{eq: dust_mass} and \ref{eq: gas_mass} with the dust temperature calculated using equation \ref{eq:LTE_Tkin}. The GMC properties are summarized in Table \ref{tab:GMCs}. 

To compare the dust emission at different scales, we need to extrapolate the dust flux at 220 GHz to the flux at 345 GHz. We assume the dust is still optically thin so that the grey-body dust spectrum is 
\begin{equation}
\label{eq:gray_body}
S_{\nu, \mathrm{dust}} \propto \frac{\nu^{\beta+3}}{e^{h\nu/kT}-1} \propto \nu^{\beta+2}
\end{equation}
\citep{Casey_2012} where $\beta$ is the dust emissivity index and we assume $\beta = 1.5$. The extrapolated flux at 345 GHz can be calculated as
\begin{equation}
S_{\mathrm{345GHz, dust}} = S_{\mathrm{220 GHz, dust}} \left(\frac{345 \mathrm{GHz}}{220 \mathrm{GHz}}\right)^{3.5}
\end{equation}

The flux ratios are shown in the left panel of Fig. \ref{fig:flux_ratio}. As we can see, the flux ratio for free-free emission is about 50\% while the dust emission fraction is $\sim$20\%. The free-free emission traces the ionized gas component, which lets us calculate the mass of the stellar component based on the SSP assumption, while the dust emission traces the gas component. Therefore, we can conclude that the stellar component is more concentrated in YMCs than is the gas component. We note that the gas in the YMCs is generally warmer than the gas outside the YMCs (Tables \ref{tab:YMCs_measured} and \ref{tab:GMCs}). As indicated by Eq. \ref{eq:gray_body}, warmer dust is more luminous. Hence the fraction of central gas over total GMC gas mass may be even lower than the observed luminosity ratio of 20 \%. These results are consistent with simulations \citep{Li_2019} that show that the stellar component is more radially concentrated than the gas component in GMCs. 

Fig. \ref{fig:flux_ratio} also shows that sources 3 and 4 have flux ratios greater than 100\%. If we increase the sensitivity of the image by using a robust parameter of 2.0 instead of 0.5, we recover even higher fluxes at YMC scales that exceed the allowed uncertainty range. Since our GMC-scale map has a LAS of $\sim$70", this difference is unlikely to be caused by missing flux at large scales. Because the YMC-resolution data were taken 2 years before the GMC-resolution data, it is possible that these two sources are time-variable sources with decreasing luminosity with time. As shown in Fig. \ref{fig:cont_image}, these two sources are close to the southern nucleus and so it is possible these sources are AGN or supernovae.

\subsection{Cluster Formation Efficiency}

The ratio of free-free emission at YMC scales compared to GMC scales basically characterizes what fraction of stars are formed in bound star clusters, which is equivalent to the cluster formation efficiency (CFE). We ignore the "cruel cradle factor" \citep{Kruijssen_2012}, which is the fraction of stars that remain bound after the cloud is dispersed. \citet{Kruijssen_2012} derived theoretical predictions for the global CFE of galaxies. Among all the factors, CFEs are most strongly dependent on the mean gas surface density of the galaxy. To compare with this theoretical prediction, we plot our free-free flux ratio versus the GMC surface density in the right panel of Fig. \ref{fig:flux_ratio}. The GMC surface density is derived from the dust flux at 220 GHz continuum. We also overlay the literature data \citep[compiled by][]{Kruijssen_2012,Adamo_2020} for comparison; in these studies, each data point represents a measurement for a single galaxy. We can see our measurements for individual GMCs are close to but below the theoretical prediction. Since our data points are all at high gas surface density end, we cannot use our new data on its own to fit the whole trend and see if it agrees with the theoretical prediction. 

We note that in the model of \citet{Kruijssen_2012}, both CFE and gas surface density are global quantities for each individual galaxy. Here we are trying to apply this model to GMC scales. The major effect is that the model is averaging gas at much larger scales and so the gas surface density should be lower than the surface density for individual GMCs, which could bring the data points leftward.

\subsection{Mass correlation between GMCs and YMCs}

\begin{figure*} 
	\plottwo{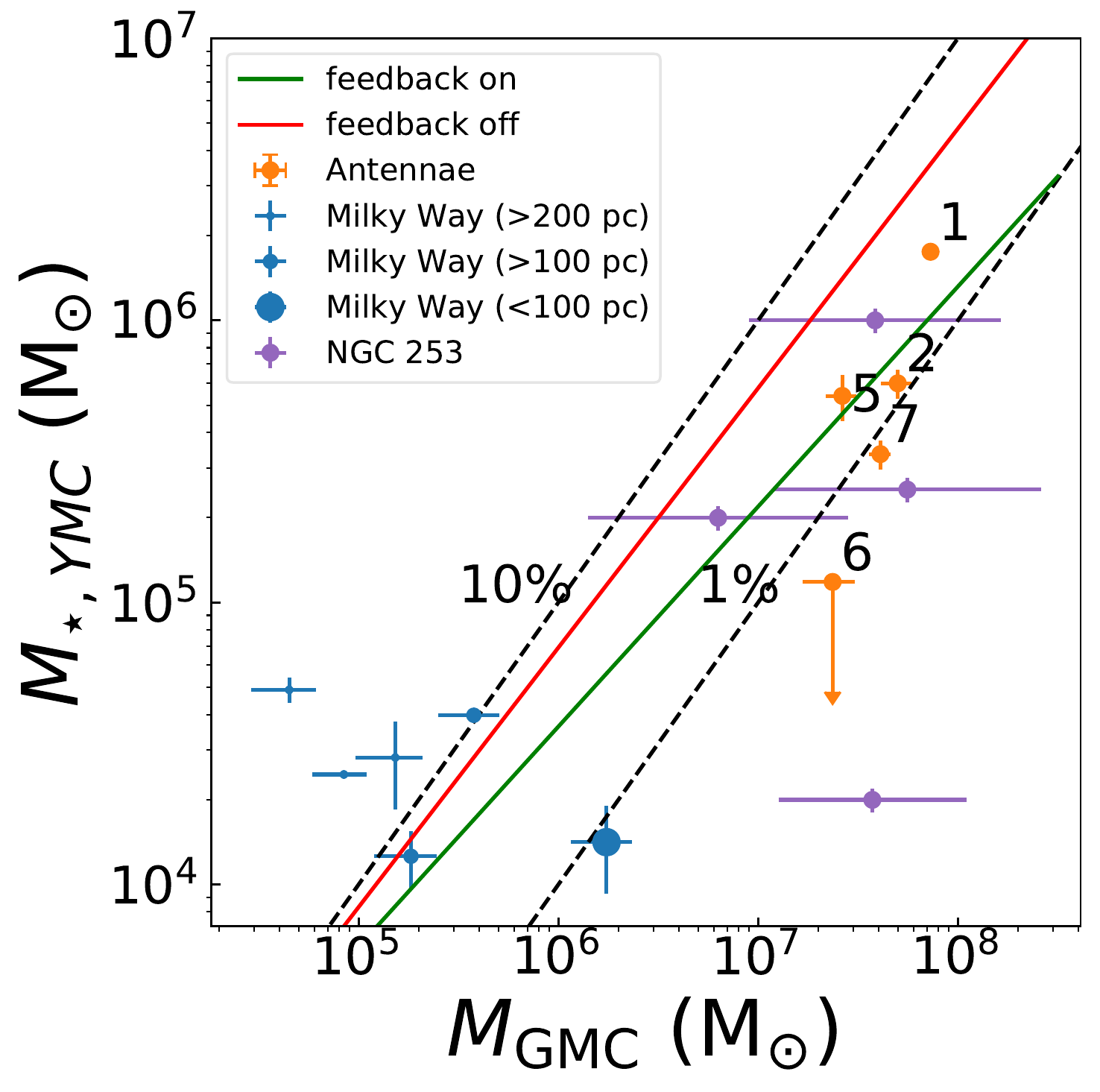}{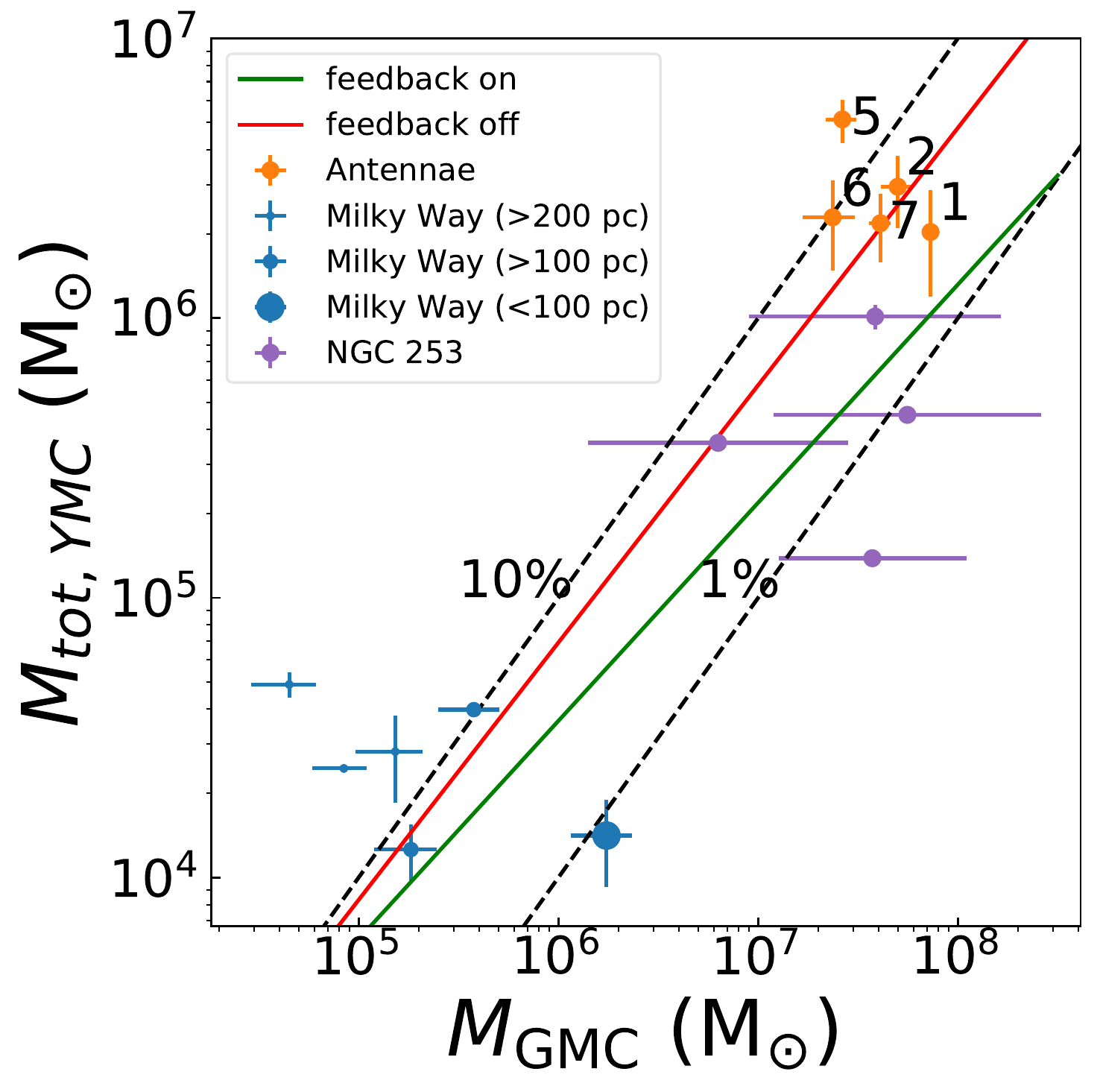}
	\caption{YMC stellar mass (Left) and total mass (Right)  versus the host GMC mass. The red and green solid lines are simulations from \citet{Howard_2018} with feedback off and on. The two dashed lines are constant ratios of 0.1 and 0.01. We can see the observational data generally agree with the simulation predictions. }
	\label{fig: MGMC_MYMC}
\end{figure*}

Clusters generally form in the densest parts of GMCs \citep{Krumholz_2019}. As they form, feedback starts to take effect and eventually removes the gas, thereby limiting the star formation efficiency (SFE). It is known from observations that the mass functions of YMCs and GMCs have similar power-law slopes of -2.0, which suggests a relatively constant SFE \citep[e.g.][]{Mok_2020}. However, the statistical cutoff due to the rarity of GMCs and YMCs at the upper mass end  makes it hard to study this relation for the massive clouds ($\sim 10^8$ \solarmass) that are common in LIRGs and ULIRGs. On the other hand, idealized simulations confirm a tight correlation between the maximal cluster mass and GMC mass \citep{Howard_2018} up to GMC masses of 10$^6$ \solarmass. Radio observations can probe the most massive YMCs ($\sim 10^6$ \solarmass) when they are still associated with clouds, thus enabling us to match these YMCs with their host GMCs.

In Fig. \ref{fig: MGMC_MYMC}, we plot the stellar mass and total mass of YMCs versus GMC molecular gas mass for the Antennae, NGC 253 and Milky Way. For the Milky Way, the stellar masses of YMCs are from \citet{Krumholz_2019}. Since most of these YMCs are already outside the host molecular cloud, we assume the total masses of these clusters are the same as their stellar masses. 
We then match the Milky Way YMCs with their closest GMCs using the GMC catalog in \citet{Rice_2016}. We excluded matched clouds with a closest distance greater than 300 pc. If there are more than two clusters belong to the same cloud, we only show the cluster with the maximum stellar mass. For NGC 253, the YMC data is from \citet{Leroy_2018} while the GMC data is from \citet{Leroy_2015}. We overlay the YMCs on Fig. 9 in \citet{Leroy_2015} to spatially match each YMCs with its corresponding GMC. We then compare the central velocity of the YMCs with their matched GMCs to confirm this correspondence. If the central velocity does not match, we find the closest cloud that has a velocity range covering the YMC central velocity. The same rule that we only show the cluster with maximum stellar or total mass applies if we have more than 1 YMC corresponding a single GMC. We also plot the simulation predictions for the YMC-GMC mass relation from \citet{Howard_2018} with feedback on and off. Note that this relation is between the GMC mass and maximum YMC mass within that GMC. The upper limit of the GMC mass in this simulation is $10^6$ \solarmass and so we must extrapolate the relation to higher masses of $\sim 10^8$ \solarmass. 

From Fig. \ref{fig: MGMC_MYMC}, we can see the data generally agrees with the correlations predicted by the simulation. For the Milky Way data points, we use symbols of different sizes to represent clusters within different distance ranges away from their matched GMCs. We can see that YMCs more than 200 pc away from their matched GMCs are all far above the simulation predictions. Since GMC diameters are generally smaller than 100 pc in Milky Way \citep{Heyer_2015}, we would expect those YMCs are outside their matched GMCs. A probable scenario is that those YMCs have already dispersed a significant amount of gas in their host GMCs and hence their matched GMC mass is less than what we expect. In contrast, the YMC with distance less than 100 pc from its matched GMC shows a stellar masses less than what we expect. This YMC is probably still embedded in its matched GMC and may be continuing to form stars.  

We can also see that YMCs in NGC 253 have a similar mass range as YMCs in the Antennae. For NGC 253, there is one data point significantly below the simulation predictions. We would expect that YMC is very young and lots of stars are yet to form. In the right panel of Fig. \ref{fig: MGMC_MYMC} where we plot the total YMC mass versus the GMC gas mass, we can see that data point is closer to the simulation predictions. YMCs in the Antennae show less scatter in the left panel and seem to agree better with the feedback-on relation from \citet{Howard_2018}. However, we note that those values may be lower limits since some of the YMCs are still going to form stars (see Section 4.1). If we plot the total mass versus the GMC mass, we can see those points are clustered around the feedback-off relation. However, as we discussed in Section 4.2, we probably include a lot of gas that is outside of the YMCs. Therefore, the data points in the right panel of Fig. \ref{fig: MGMC_MYMC} should be considered as upper limits. Also, due to our limited sample size and heterogeneous data sources, the results in this section should be considered to be illustrative, still in need of further investigation. 

\subsection{No Effect of YMCs on GMC temperatures}

\begin{figure}
	\plotone{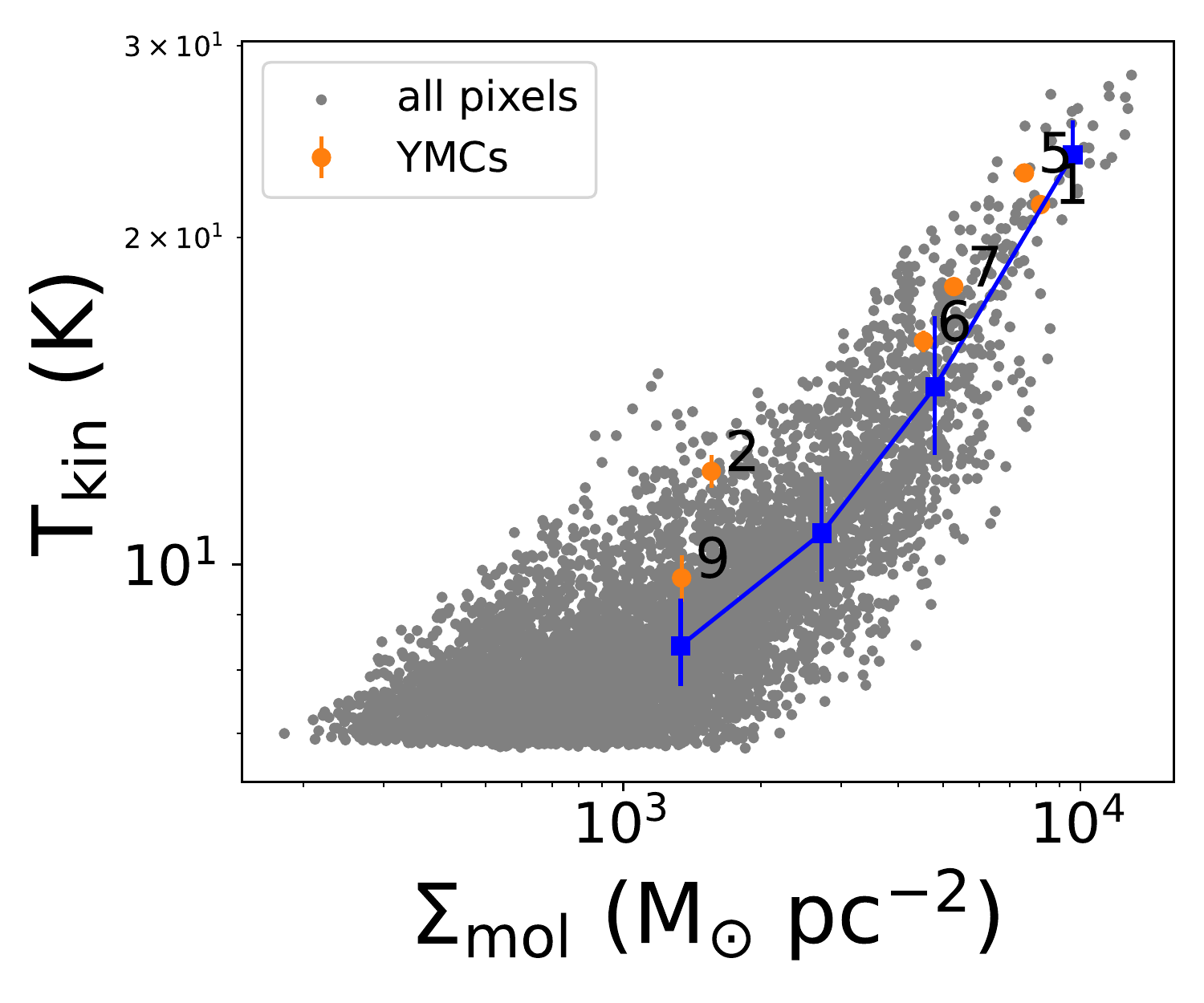}
	\caption{The GMC temperature derived from CO 2-1 observations at resolution of $\sim$60 pc versus the molecular gas surface density. We assume a ULIRG \alphaco of 1.1 \alphacou. The gray points are all pixels detected in the \cotwo map. The orange points are pixels in the \cotwo map that coincide with peaks of the continuum source. The blue points and error bars represent the median and 25 and 75 percentile value at each given bin for the gas surface density. There is no strong evidence that the temperatures of GMCs that are forming YMCs are significantly higher than the temperatures of the remaining GMCs.}
	\label{fig:LTE_Tkin}
\end{figure}

It is interesting to see if clouds with YMCs have different properties from clouds without YMCs. As we know, YMCs have strong free-free emission, which could heat the dust and make the clouds warmer. Therefore, we might expect a temperature difference between clouds with or without YMCs. We apply equation \ref{eq:LTE_Tkin} to the \cotwo data at GMC resolution (Brunetti et al. in prep). To avoid effects from correlated pixels, we Nyquist sample the image by rebinning the pixels to half of the beam. Fig. \ref{fig:LTE_Tkin} shows the calculated $T_{\mathrm{kin}}$ versus the surface density of the molecular gas. Note that in this plot, the gas surface density is calculated based on the GMC-resolution \cotwo cube since we do not have 220 GHz continuum detections for all \cotwo detected pixels. The gray points show all the detected pixels in the \cotwo map with peak brightness temperature greater than 10 times the rms noise. The orange points are the peak value of the pixels within our apertures used to measure the flux of the continuum point source at GMC scales. Those points represent the properties of the GMCs that host radio YMCs. We also divide the data points into different bins based on the gas surface density values and calculate the median for each bin (blue points; error bars show the value of 1st and 3rd quartile of each bin). 

From the plot, we can see the orange points are generally above the blue points. However, at the high surface end (5000 -- 10000 \coldenunit), 3 of 4 YMCs have temperatures lower than the first quartile values, which indicates their temperature is not significantly different from the rest of pixels. On the other hand, we see the 2 orange points are above the first quartile values at the low surface density end ($\sim 1000$ \coldenunit). However, in this surface density regime, we are hitting the sensitivity limit of the 100 GHz YMC-resolution continuum. Source 9 is barely detected in our high-resolution continuum image so we can consider it to mark the lower detection limit. Therefore, our methods to identify YMCs bias towards sources with high peak brightness temperatures at low mass surface densities. Overall, we see no clear evidence that the feedback from YMCs has increased the temperature of the host GMCs yet. However, 
since we only have a limited number of radio sources, our results cannot conclusively show whether YMC feedback has affected GMCs yet. In the future, we will consider to add young optical star clusters to this type of analysis. 

\section{Conclusions}

In this paper, we have presented new, high-resolution continuum data for  YMCs in the Antennae. We combine these data with CO and continuum data at GMC scales to explore the correlation of properties between the  YMCs and their host GMCs. Our main conclusions are summarized below. 

\begin{itemize}
	\item These YMCs have stellar masses of $\sim$ $10^5$--$10^6$ \solarmass and radius of $\sim$ 3--15 pc. For sources 1b, 2 and 5, we can observe substructures in higher-resolution images, which suggest we might overestimate their radii by including some surrounding diffuse emission. Source 1a and 7 still look compact at the highest resolution.

	\item Based on statistical counts, we estimate the lifespan of these YMCs to be about 1 Myr. This is consistent with estimates of embedded YMC lifetimes. However, all these sources have Pa$\beta$ counterparts. By comparing the coordinates and fluxes between 100 GHz and Pa$\beta$ sources, we think these YMCs have partly emerged and have already ionized some diffuse medium outside the cloud.  
	
	\item A virial analysis of these YMCs suggests the majority are bound systems. This further suggests that they may appear as young globular clusters after the gas is dispersed.
	  
	\item More than 50\% of the free-free emission at GMC scales comes from compact YMCs inside those GMCs. This fraction is equivalent to the cluster formation efficiency (CFE). We compare this fraction with the theoretical prediction from \citet{Kruijssen_2012} and literature data for galaxies from \citet{Adamo_2020} and see a quite good agreement. 
	
	\item We explore the correlation between the YMC mass and its host GMC mass. We also compare this correlation in the Antennae with those in NGC 253 and the Milky Way. We find the data generally agree with the  predicted correlation from simulations \citep{Howard_2018}.
	
	\item When comparing the gas temperature in regions with and without  YMCs, we find no significant difference between those two populations of clouds. We see no clear evidence that YMC feedback has increased the cloud temperature at GMC scales.  
\end{itemize}

\noindent We thank the referee for thoughtful comments and constructive suggestions. This paper makes use of the following ALMA data: \\
ADS/JAO.ALMA \#2016.1.00041.S \\
ADS/JAO.ALMA \#2016.1.00924.S \\
ADS/JAO.ALMA \#2018.1.00272.S. ALMA is a partnership of ESO (representing its member states), NSF (USA) and NINS (Japan), together with NRC (Canada), MOST and ASIAA (Taiwan), and KASI (Republic of Korea), in cooperation with the Republic of Chile. The Joint ALMA Observatory is operated by ESO, AUI/NRAO and NAOJ. The National Radio Astronomy Observatory is a facility of the National Science Foundation operated under cooperative agreement by Associated Universities, Inc. This research has made use of NASA's Astrophysics Data System Bibliographic Services. This research made use of Astropy,\footnote{http://www.astropy.org} a community-developed core Python package for Astronomy \citep{astropy:2013, astropy:2018}. The research of C.D.W. is supported by grants from the Natural Sciences and Engineering Research Council of Canada and the Canada Research Chairs program. The research of H.H. is partially supported by the New Technologies for Canadian Observatories, an NSERC-CREATE training program. The research of K.J. is supported by NSF grant 1716335.

\vspace{5mm}
\facilities{ALMA}
\software{astropy \citep{astropy:2013,astropy:2018}, 
		  CASA \citep{McMullin_2007}
          }

\bibliography{references}{}
\bibliographystyle{aasjournal}

\end{document}